

\input harvmac
\def\rhob{{\rho\kern-0.465em \rho}}

\def\ontopss#1#2#3#4{\raise#4ex \hbox{#1}\mkern-#3mu {#2}}
\def\rh{\hat{r}}
\def\sh{\hat{s}}

\setbox\strutbox=\hbox{\vrule height12pt depth5pt width0pt}

\def\strut{\relax\ifmmode\copy\strutbox\else\unhcopy\strutbox\fi}

\nref\rgrs{B. Gato--Rivera and A.M. Semikhatov, Phys. Lett. B293
(1992) 72}
\nref\rblnwa{M. Bershadsky, W. Lerche, D. Nemeschansky and N.P. Warner,
Phys. Lett. B292 (1992) 35.}
\nref\rblnwb{M. Bershadsky, W. Lerche, D. Nemeschansky and N.P. Warner,
Nucl. Phys. B401 (1993) 304.}
\nref\rabf{G.E. Andrews, R.J. Baxter and P.J. Forrester, J. Stat. Phys. 35
(1984) 193}
\nref\rdjmo{E. Date, M. Jimbo, T. Miwa, and M. Okado, Phys. Rev. B 35
(1987) 2105.}
\nref\rbail{W.N. Bailey, Proc. Lond. Math. Soc (2) 50 (1949) 1.}
\nref\rroga{L.J. Rogers, Proc. Lond. Math. Soc. 25 (1894) 318.}
\nref\rrogb{L.J..Rogers, Proc. Lond. Math. Soc. (2) 16 (1917) 315.}
\nref\rslata{L.J. Slater, Proc. Lond. Math. Soc. (2) 52 (1951) 460.}
\nref\rslatb{L.J. Slater, Proc. Lond. Math. Soc. (2) 54 (1951-52)
147.}
\nref\rand{G.E. Andrews, Pac. J. Math. 114,(1984) 276.}
\nref\raab{A.K. Agarwal, G.E. Andrews and D.M. Bressoud, J. Indian
Math. Soc. 51 (1987) 57.}
\nref\rml{S.C. Milne and G. M. Lilly, Bull. Am. Math. Soc. (NS) 26
(1992) 258.}
\nref\rfqa{O. Foda and Y--H. Quano, hepth 9408086.}
\nref\rmel{E. Melzer, Int. J. Mod. Phys. A 9 (1994) 1115.}
\nref\rber{A. Berkovich, Nucl. Phys. B431 (1994) 315.}
\nref\rfqb{O. Foda and Y--H. Quano, hepth 9407191.}
\nref\rmccoy{R. Kedem, T.R Klassen, B.M. McCoy and E. Melzer,
Phys. Letts. B 307 (1993) 68.}
\nref\rwara{ S.O. Warnaar, J. Stat. Phys. (in press) hepth 9501134.}
\nref\rwarb{ S.O. Warnaar, J. Stat. Phys. (in press) hepth 9508079.}
\nref\ranne{A. Schilling, Nucl..Phys.B (in press) hepth 9508050 and 9510168.}
\nref\razam{Al. Zamolodchikov, Nucl. Phys. B358 (1991) 524.}
\nref\rfqr{G. Feverati, E. Quattrini and F. Ravinini, hepth 9412104.}
\nref\rbm{A. Berkovich and B.M. McCoy, Letts. Math. Phys. (in press)
hepth 9412030.}
\nref\rbms{A. Berkovich, B.M.McCoy and A. Schilling, (in preparation)}
\nref\rbailb{W.N. Bailey, Quart. J. Math. (Oxford), 7 (1936) 105.}
\nref\randb{G.E. Andrews, in {\it The Theory and Application of Special
Functions} (R. Askey, ed.) Academic Press, New York (1975) 191.}
\nref\rrc{A. Rocha-Caridi, in {\it Vertex Operators in Mathematics
and Physics}. ed. J. Lepowsky, S. Mandelstam and I.M. Singer
(Springer, Berlin 1985).}
\nref\rgko{P. Goddard, A. Kent and D. Olive, Comm. Math. Phys. 103
(198) 105.}
\nref\rmat{Y. Matsuo, Prog. Theo. Phys. 77 (1987) 793.}
\nref\rdob{V.K. Dobrev, Phys. Letts. B186 (1987) 43.}
\nref\rry{F. Ravanini and S--K. Yang, Phys. Letts. B (1987) 202.}
\nref\rkir{E.B. Kiritsis, Int. J. Mod. Phys. A3 (1988) 1871.}
\nref\rbaver{E. Baver and D. Gepner, hepth 9502118.}

\nref\rfb{P.J. Forrester and R.J. Baxter, J. Stat. Phys. 38 (1985)
435.}
\nref\rzamfat{A.B. Zamolochikov and V.A. Fateev, Sov. Phys. JETP 63
(1986) 913.}
\nref\rahn{C. Ahn,Phys. Letts. B 294 (1992) 204.}
\Title{\vbox{\baselineskip12pt\hbox{BONN-TH-95-24}
  \hbox{ITPSB 95-57}
  \hbox{HEP-TH 95XXXXX}}}
  {\vbox{\centerline{$N=2$ Supersymmetry and Bailey Pairs}}}
  \centerline{Alexander Berkovich~\foot{berkovic@axpib.physik.uni-bonn.de}}

  \bigskip\centerline{\it Physikalisches Institut der}
  \centerline{\it Rheinischen Friedrich-Wilhelms Universit{\"a}t Bonn}
  \centerline{\it Nussallee 12}
  \centerline{\it D-53115 Bonn, Germany}
  \bigskip
  \centerline{and}
  \bigskip
  \centerline{ Barry~M.~McCoy~\foot{mccoy@max.physics.sunysb.edu} and
               Anne Schilling\foot{anne@insti.physics.sunysb.edu}}

  \bigskip\centerline{\it Institute for Theoretical Physics}
  \centerline{\it State University of New York}
  \centerline{\it Stony Brook,  NY 11794-3840}
  \bigskip
  \Date{\hfill 12/95}

  \eject

\centerline{\bf Abstract}

We demonstrate that the Bailey pair formulation of Rogers-Ramanujan
identities unifies the calculations of the characters  of $N=1$
and $N=2$ supersymmetric conformal field theories with the counterpart
theory with no supersymmetry. We  illustrate  this construction for
the $M(3,4)$ (Ising) model where the Bailey pairs have been given by
Slater. We then present the general unitary case. We  demonstrate that
the model $M(p,p+1)$ is derived from $M(p-1,p)$ by a Bailey
renormalization flow and conclude by
obtaining the $N=1$ model $SM(p,p+2)$ and the unitary $N=2$ model
with central charge $c=3(1-2/p).$

\newsec{Introduction}

In the past several years it has been observed~\rgrs--\rblnwb~
that string theories
of two dimensional quantum gravity have a strong connection with
$N=2$ superconformal models. This observation was first made in~\rgrs~ and
is forcefully presented in~\rblnwb~ where it is stated ``The underlying
$N=2$ superconformal symmetry is quite generic and is present in every
string theory.''

It has also been realized since the work of Andrews, Baxter, and
Forrester~\rabf~and Date, Jimbo, Miwa and Okado~\rdjmo~that there
is a thoroughgoing relationship between
Rogers--Ramanujan identities and conformal field theory
characters. Consequently the relation seen between $N=2$ superconformal
field theories and string theory should  be
expected to be present in the theory of Rogers--Ramanujan
identities. It is the purpose of this paper to demonstrate that this
is correct and that the characters of $N=2$ superconformal models
may be obtained from the nonsupersymmetric models $M(p,p')$ by means
of the construction known as the Bailey pair. Indeed it is our
contention that this relation between string theory and Bailey pairs
is much more than an analogue and that the Bailey construction
provides an exact reformulation of the string theory in terms of a
completely fermionic description of the Fock spaces.

The method of the
 Bailey pair
was invented by Bailey~\rbail~in his proofs of many of the
results of Rogers~\rroga--\rrogb. This method was extensively used by
Slater~\rslata--\rslatb, has been extended by Andrews~\rand, was
extended to the Bailey lattice by Agarwal, Andrews and
Bressoud~\raab~, and generalized to higher rank groups  by Milne
and Lilly~\rml. Most recently
this technique has been combined by Foda and Quano ~\rfqa~with the
finite size results on $M(p,p+1)$ of \rmel--\rber~and $M(2,2k+1)$ of
{}~\rfqb~to provide proofs of various fermionic characters formulas
conjectured in~\rmccoy.

By definition a pair of sequences ($\alpha_n,~\beta_n$) is said to form
a Bailey pair (for the group $A_1$ relative to the parameter $a$)
if
\eqn\bail{\beta_n=\sum_{j=0}^{n}{\alpha_j\over (q)_{n-j}(aq)_{n+j}}}
where we use the definition
\eqn\isingb{(a;q)_n=\prod_{l=0}^{n-1}(1-aq^l)}
which, when there is no danger of confusion, we abbreviate as $(a)_n$.
The inverse of this relation is
\eqn\bailinv{\alpha_n={1-aq^{2n}\over
1-a}\sum_{j=0}^n{(a)_{n+j}(-1)^{n-j}q^{{1\over 2}(n-j)(n-j-1)}\over
(q)_{n-j}}\beta_j.}

The lemma of Bailey states that whenever $(\alpha_n,~\beta_n)$ is
a Bailey pair related by~\bail~then
\eqn\lemma{\eqalign{\sum_{n=0}^{N}&{(\rho_1)_n (\rho_2)_n(aq/\rho_1
\rho_2)_{N-n}(aq/\rho_1 \rho_2)^n\beta_n\over (q)_{N-n}(aq/\rho_1)_N
(aq/\rho_2)_N}\cr
=\sum_{n=0}^{N}&\biggl({(\rho_1)_n
(\rho_2)_n(aq/\rho_1
\rho_2)^n \alpha_n\over (aq/\rho_1)_n (aq/\rho_2)_n}\biggr)
{1\over (q)_{N-n}(aq)_{N+n}}.\cr}}
The proofs of these results are given in ~\rbail~and ~\rand.

The lemma of Bailey is a result for rational functions. By taking
the limit $N\rightarrow \infty$ we
obtain from ~\lemma~ the theorem that if
$(\alpha_n,\beta_n)$ is a Bailey pair then
\eqn\inflemma {\sum_{n=0}^{\infty}
(\rho_1)_n(\rho_2)_n(aq/\rho_1\rho_2)^n\beta_n\
={(aq/\rho_1)_{\infty}(aq/\rho_2)_{\infty}\over
(aq)_{\infty}(aq/\rho_1\rho_2)_{\infty}}\sum_{n=0}^{\infty}\biggl({(\rho_1)_n
(\rho_2)_n(aq/\rho_1
\rho_2)^n \alpha_n\over (aq/\rho_1)_n (aq/\rho_2)_n}\biggr).}
This result has been extensively used by Slater~\rslata--\rslatb~to
prove many of the identities first derived by Rogers~\rroga--\rrogb~by
making the following two specializations of the parameters $\rho_1,~\rho_2;$
\eqn\speca{1:~~~\rho_1\rightarrow\infty,~~~\rho_2\rightarrow\infty}
\eqn\specb{2:~~~\rho_1\rightarrow\infty,~~~\rho_2={\rm finite},}
and from her extensive list of results~\rslatb~one finds the
remarkable result that the first specialization~\speca~leads to characters
of minimal models $M(p,p')$ and the second specialization
{}~\specb~leads to characters of $N=1$ supersymmetric models $SM(p,p').$

There is, however, a third case not considered by Slater
\eqn\specc{3:~~~\rho_1={\rm finite},~~\rho_2={\rm finite}.}
We show here  that this leads to characters
of the $N=2$ supersymmetric models.

It cannot, of course, be an accident  that Bailey's
construction, invented decades before conformal field theory was even
thought about, gives characters of the $N=0,~ 1$ and $2$
supersymmetric conformal field theories in a unified manner.  In
fact there is a complete connection between the Bailey construction,
the theory of affine Lie algebras and even two dimensional quantum
gravity and this paper is meant to be the first of several in which we
present these topics in detail. However, there are several technical
complications which must be presented in some detail in order to deal
with nonunitary models which tend to obscure the emergence of the
$N=2$ supersymmetry. Consequently in this first paper we will
introduce the subject by restricting our attention to the unitary
models $M(p,p+1)$ and their $N=1$ and $N=2$ supersymmetric extensions.

We begin in sec. 2 by reviewing the construction of
Slater~\rslata--\rslatb~of the characters of $M(3,4)$ and $SM(3,5)$ and
then demonstrate how the characters of the $N=2$ unitary model with
$c=1$ are obtained by using the specialization~\specc. In sec. 3 we
present a second construction of Slater~\rslata~which gives $M(3,4)$
under the specialization ~\speca~but gives a special case of the $N=2$
model under ~\specb. We finally consider the general unitary model
$M(p,p+1)$ in sec. 4 where we first obtain the Bailey pairs from the
finite size polynomial computations of ~\rabf,~\rmel,
{}~\rber,~\rwara--\ranne~by use of the
method of ~\rfqa.~We obtain all the characters of $M(p,p+1)$ from this
construction and thus extend the results of~\rfqa~and~\rwarb.
The relation of this construction, which we call Bailey
renormalization flow, to
the renormalization flows of~\razam~(and~\rfqr)~is
first proposed in~\rbm.
We
then obtain from the Bailey pairs the $N=1$ models
$SM(p,p+2)$
and the unitary
$N=2$ models with central charge $c=3(1-2/p).$ We  conclude in sec. 5 with a
few remarks
on the more general case which will be presented more fully elsewhere~\rbms.

\newsec{The Ising model $M(3,4)$ and its $N=1$ and $N=2$ extensions.}

The Ising model is the minimal model $M(3,4).$ It has three
independent characters which are obtained from the general bosonic formula for
the characters of the $M(p,p')$ model
\eqn\rocca{\chi_{r,s}^{(p,p')}(q)=\chi_{p-r,p'-s}^{(p,p')}(q)={1\over
(q)_{\infty}}\sum_{j=-\infty}^{\infty}
\bigl(q^{j(jpp'+rp'-sp)}-q^{(jp'+s)(jp+r)}\bigr).}
Here $p$ and $p'$ are relatively prime, the central charge is
\eqn\cen{c=1-{6(p-p')^2\over pp'}}
and the conformal dimensions are
\eqn\dim{\Delta_{r,s}^{(p,p')}={(rp'-sp)^2-(p-p')^2\over 4pp'}~~(1\leq
r\leq p-1,~1\leq s \leq p'-1).}

These characters also have a fermionic representation as $q$ series
due to the following  identities proven in 1894 by Rogers~\rroga~
\eqn\isinga{\sum_{m=0,{\rm even}}^{\infty}{q^{m^2/ 2}\over
(q)_m}=\chi_{1,1}^{(3,4)}(q),~~~~~~~
\sum_{m=1,{\rm odd}}^{\infty}{q^{m^2/ 2}\over
(q)_m}=\chi_{1,3}^{(3,4)}(q),}
\eqn\isingaa{\sum_{m=0, {\rm even}}^{\infty}{q^{m(m-1)/ 2}\over (q)_m}=
\sum_{m=1, {\rm odd}}^{\infty}{q^{m(m-1)/2}\over (q)_m}
=\chi_{1,2}^{(3,4)}(q).}

Slater~\rslata~ obtains the four identities~\isinga~\isingaa~from the following
four
Bailey pairs (which following the notation of Rogers she calls
A(5)-A(8)) where $\alpha_0=1$ in all cases and


\vbox{\tabskip=0pt

\offinterlineskip
\halign to 125pt{\strut# \quad &\hfil #\quad&\hfil #\quad&\hfil #\quad&\hfil
#\quad & \hfil #\quad&
\hfil #\quad&# \cr

&$~$&$\beta_n$&$\alpha_{3n-1}$&$\alpha_{3n}$&$\alpha_{3n+1}$&$a$&\cr
&A(5)&$q^{n^2}/(q)_{2n}$&$-q^{3n^2-n}$&$q^{3n^2-n}+q^{3n^2+n}$&$-q^{3n^2+n}$
&$1$&\cr
&A(6)&$q^{n^2}/(q^2)_{2n}$&$q^{3n^2+n}$&$q^{3n^2-n}$&$-q^{3n^2+n}-q^{3n^2+5n+2}$
&$q$&\cr
&A(7)&$q^{n^2-n}/(q)_{2n}$&$-q^{3n^2-4n+1}$&$q^{3n^2-2n}+q^{3n^2+2n}$
&$-q^{3n^2+4n+1}$&$1$&\cr
&A(8)&$q^{n^2+n}/(q^2)_{2n}$&$q^{3n^2-2n}$&$q^{3n^2+2n}$&
$-q^{3n^2+4n+1}-q^{3n^2+2n}$&$q$&\cr}}
When we put these four pairs into~\inflemma~we find from A(5) and
A(8)
 that for $a=1,q$ (after a bit of algebra)
\eqn\isingc{\eqalign{\sum_{n=0}^{\infty}(\rho_1)_n(\rho_2)_n&
(aq/\rho_1\rho_2)^n
{q^{n^2}a^n\over(aq)_{2n}}
={(aq/\rho_1)_{\infty}(aq/\rho_2)_{\infty}\over
(aq)_{\infty}(aq/\rho_1\rho_2)_{\infty}}
\sum_{j=-\infty}^{\infty}\cr
&\times\biggl({(\rho_1)_{3j}(\rho_2)_{3j}
(aq/\rho_1\rho_2)^{3j}\over
(aq/\rho_1)_{3j}(aq/\rho_2)_{3j}}-
{(\rho_1)_{3j+1}(\rho_2)_{3j+1}(aq/\rho_1\rho_2)^{3j+1}\over
(aq/\rho_1)_{3j+1}(aq/\rho_2)_{3j+1}}\biggr)a^j q^{3j^2+j}\cr}}
where we define
\eqn\min{(a)_{-n}={1\over (1-aq^{-1})(1-aq^{-2})\ldots(1-aq^{-n})}
 ={1\over (aq^{-n})_n}={(-q/a)^nq^{{1\over 2}n(n-1)}\over (q/a)_n}.}
Similarly we find from A(6) and A(7) that with $a=1,q$
\eqn\isingd{\eqalign{
\sum_{n=-\infty}^{\infty}(\rho_1)_n(\rho_2)_n&
(aq/\rho_1\rho_2)^n
{q^{n(n-1)}a^n\over(aq)_{2n}}
={(aq/\rho_1)_{\infty}(aq/\rho_2)_{\infty}\over
(aq)_{\infty}(aq/\rho_1\rho_2)_{\infty}}
\sum_{j=-\infty}^{\infty}a^jq^{3j^2-2j}\cr
&\times\biggl({(\rho_1)_{3j}(\rho_2)_{3j}
(aq/\rho_1\rho_2)^{3j}\over
(aq/\rho_1)_{3j}(aq/\rho_2)_{3j}}-{(\rho_1)_{3j-2}(\rho_2)_{3j-2}
(aq/\rho_1\rho_2)^{3j-2}\over
(aq/\rho_1)_{3j-2}(aq/\rho_2)_{3j-2}}\biggr)\cr}}
These two identities hold for all values of $\rho_1$ and $\rho_2.$
By making suitable specializations of these parameters these formulas
specialize  to the  characters of several different conformal field
theory models. There are three distinct cases and we will treat them
separately.

\subsec{The model $M(3,4)$}

Slater~\rslatb~obtains the Ising model identities~\isinga~\isingaa~
from~\isingc~and
{}~\isingd~by making the specialization~\speca~
where, using
\eqn\inflim{\lim_{\rho\rightarrow\infty}\rho^{-n}(\rho)_n
=(-1)^nq^{{n\over 2}(n-1)}}
we find from~\isingc~
\eqn\isingf{\sum_{n=0}^{\infty}{q^{2n^2}a^{2n}\over (aq)_{2n}}={1\over
(aq)_{\infty}}\sum_{j=-\infty}^{\infty}\biggl(q^{j(12j+1)}
a^{4j}-q^{(3j+1)(4j+1)}a^{4j+1}\biggr).}
Then, by setting $a=1$ and $a=q$ we obtain the identities of
{}~\isinga~for $\chi_{1,1}^{(3,4)}(q)$ and $\chi_{1,3}^{(3,4)}(q).$
These are (83) and (86) of ~\rslatb.
Similarly from~\isingd~ we find
\eqn\isingg{\sum_{n=0}^{\infty}{q^{2n^2-n}a^{2n}\over
(aq)_{2n}}={1\over (aq)_{\infty}}\sum_{j=-\infty}^{\infty}
\biggl(q^{12j^2-2j}a^{4j}-q^{(3j+1)(4j+2)}a^{4j+2}\biggr).}
and by setting
$a=1$ and $a=q$ we find
the two forms of the identity~\isingaa~ for $\chi_{1,2}^{(3,4)}(q).$ These are
(84) and (85) of ~\rslatb.

\subsec{The $N=1$ supersymmetric model $SM(3,5)$}

The second set of specializations made by Slater~\rslatb~is
\eqn\spect{\rho_1\rightarrow\infty,~~~\rho_2=-q^{1\over 2}~{\rm and}~-q.}
We will see that these give characters of the $N=1$ superconformal
model $SM(p,p')$ where
\eqn\smin{{\hat \chi}_{r,s}^{(p,p')}(q)={\hat
\chi}_{p-r,p'-s}^{(p,p')}(q)=
{(-q^{\epsilon_{r-s}})_{\infty}\over
(q)_{\infty}}\sum_{j=-\infty}^{\infty}\bigl(q^{j(jpp'+rp'-sp)\over
2}-q^{(jp-r)(jp'-s)\over 2}\bigr)}
where $1\leq r\leq p-1,~1\leq s \leq p'-1,$ $p$ and $(p'-p)/2$ are
relatively prime and
\eqn\epsd{\epsilon_a=\cases{{1\over 2}&if $a$ is even (Neveu-Schwarz
(NS) sector)\cr
1& if $a$ is odd (Ramond (R) sector).\cr}}
Here the central charge is
\eqn\censs{c={3\over 2}-{3(p-p')^2\over pp'}}
 and the conformal
dimensions are
\eqn\dimss{{(rp'-sp)^2-(p-p')^2\over 8pp'}+{2\epsilon_{r-s}\over 16}}

In the first case where~$\rho_2=-q^{1\over 2}$ we consider ~\isingc~with $a=1$
and find
\eqn\isingh{\sum_{n=0}^{\infty}
{(-q^{1\over 2})_nq^{{3\over2 }n^2}\over(q)_{2n}}
={(-q^{1\over 2})_{\infty}\over
(q)_{\infty}}\sum_{n=-\infty}^{\infty}\biggl(q^{{1\over 2}n(15n-2)}
-q^{{1\over 2}(3n-1)(5n-1)}\biggr).}
which from ~\smin~is the character ${\hat \chi}_{1,1}^{3,5}(q)$ of the
model $SM(3,5)$ and is eqn.(100) in Slater's list~\rslatb.
Similarly from~\isingd~with $a=1$
\eqn\isingi{\sum_{n=0}^{\infty}(-q^{1\over 2})_n{q^{{3n^2\over 2}-n}\over
(q)_{2n}}={(-q^{1\over 2})_{\infty}\over
(q)_{\infty}}\sum_{j=-\infty}^{\infty}\biggl(q^{{1\over
2}j(15j-4)}-q^{{1\over 2}(3j+1)(5j+3)}\biggr)={\hat \chi}_{1,3}^{3,5}(q)}
which is (95) on Slater's list.

For the case where $\rho_2=-q$ we find
from~\isingc~ with $a=q$ that
\eqn\isingj{\sum_{n=0}^{\infty}(-q)_n{q^{{3n\over 2}(n+1)}\over
(q)_{2n+1}}={(-q)_{\infty}\over (q)_{\infty}}\sum_{n=-\infty}^{\infty}
\biggl(q^{{1\over 2}n(15n+7)}-q^{{1\over 2}(3n+2)(5n+1)}\biggr)=
{\hat\chi}_{2,1}^{(3,5)}(q)={\hat\chi}_{1,4}^{(3,5)}(q)}
which is (63) in Slaters list
while setting $a=q$ in ~\isingd~ we obtain
\eqn\isingk{\sum_{n=0}^{\infty}(-q)_n{q^{{n\over 2}(3n+1)}\over
(q)_{2n+1}}={(-q)_{\infty}\over (q)_{\infty}}\sum_{n=-\infty}^{\infty}
\biggl(q^{{1\over 2}n(15n+1)}-q^{{1\over 2}(3n+2)(5n+3)}\biggr)=
{\hat \chi}_{2,3}^{(3,5)}(q)={\hat \chi}_{1,2}^{(3,5)}(q)}
which is (62) of Slater.

These four identities may be put into the canonical quasi particle
form~\rmccoy~  by using the elementary identity
\eqn\isingl{(x)_n=\sum_{j=0}^{n}{n\atopwithdelims[]
j}_q(-x)^jq^{{1\over 2}j(j-1)}=
\sum_{m=0}^{n}(-x)^{(n-m)}q^{{1\over
2}(n-m)(n-m-1)}{n\atopwithdelims[] m}_q}
where the $q-$binomial coefficient is defined as
\eqn\qbin{{n\atopwithdelims[] j}_q=\cases{{(q)_n\over (q)_j
(q)_{n-j}}&for $0\leq j\leq n$\cr
0&otherwise.}}
Thus we find
\eqn\mmua{\sum_{m=0}^{\infty}
\sum_{n=0}^{\infty}{q^{2n^2+{1\over 2}m^2-nm}\over
(q)_{2n}}{n\atopwithdelims[] m}_q={\hat
\chi}_{1,1}^{3,5}(q),}
\eqn\mmub{\sum_{m=0}^{\infty}\sum_{n=0}^{\infty}
{q^{2n^2+{1\over 2}m^2-nm-n}\over
(q)_{2n}}{n\atopwithdelims[] m}_q={\hat \chi}_{1,3}^{(3,5)}(q).}
\eqn\mmug{\sum_{m=0}^{\infty}\sum_{n=0}^{\infty}{q^{2n^2+{1\over
2}m^2-nm+2n-{1\over 2}m}\over (q)_{2n+1}}{n\atopwithdelims[] m}_q=
{\hat\chi}_{1,4}^{(3,5)}(q),}
and
\eqn\mmuh{\sum_{m=0}^{\infty}\sum_{n=0}^{\infty}{q^{2n^2+{1\over
2}m^2-nm+n-{1\over 2}m}\over (q)_{2n+1}}{n\atopwithdelims[] m}_q=
{\hat\chi}_{1,2}^{(3,5)}(q)}

We have now succeeded in obtaining expressions for all four of the
independent characters of $SM(3,5).$ However just as in the case of
$M(3,4)$ the symmetry condition
${\hat\chi}_{r,s}^{(p,p')}(q)={\hat\chi}_{p-r,p'-s}^{(p,p')}(q)$
implies certain further nontrivial identities. One of these is found
by setting $a=1$ and $\rho_2=-q$ in ~\isingc~to find
a second representation for
${\hat\chi}_{1,2}^{(3,5)}(q)$ in addition to ~\isingk~of
\eqn\mmui{\sum_{n=0}^{\infty}(-q)_n{q^{{3n^2\over 2}-{n\over 2}}\over
(q)_{2n}}={\hat\chi}_{1,2}^{(3,5)}(q)}
and if we set $a=1,$ and $\rho_2=-q$ in ~\isingd~we obtain
\eqn\nnuj{\sum_{n=0}^{\infty}(-q)_n{q^{{3n\over 2}(n-1)}\over
(q)_{2n}}
={\hat\chi}_{1,2}^{(3,5)}(q)+{\hat\chi}_{1,4}^{(3,5)}(q).}

\subsec{The $N=2$ unitary supersymmetric model with $c=1$}

We now consider the third specialization~\specc. This was not
considered by Slater. We will show that
this will lead to a unitary $N=2$ supersymmetric model.
For these models the central charge is $c=3(1-2/m)$ and
 there are three sectors called A for antiperiodic
(Neveu-Schwarz), P for periodic (Ramond) and T for twisted.
In the A
and P sectors the  (normalized) characters are~\rmat-\rkir~
\eqn\ntwo{\eqalign{&\chi_{r,s}^{(N=2)(m)}(y,q)\cr
&={(-q^{\tilde
\epsilon}y)_{\infty}(-q^{\tilde \epsilon}y^{-1})_{\infty}
\over (q)_{\infty}^2}\sum_{j=-\infty}^{\infty}
q^{mj^2+(r+s)j}\biggl(1-{q^{mj+r}y^{-1}\over 1+q^{mj+r}y^{-1}}-
{q^{mj+s}y\over 1+q^{mj+s}y}\biggr)\cr}}
where:

1) In the sector A   $r$ and $s$ are half integers with
$0\leq r,s,r+s\leq m-1$ and ${\tilde
\epsilon}={1\over 2}$ and the conformal dimensions of the Virasoro
operator $L_0$ and the $U(1)$ current $J_0$ are
\eqn\ntwdima{h_{r,s}^{A}=(rs-{1\over 4})/m,~~~q_{r,s}^{A}=(r-s)/m;}

2) In the  $P$ sector $r$ and $s$ are integers with $0\leq
r-1,s,r+s\leq m-1$ and ${\tilde \epsilon}=1$ and
\eqn\ntdimb{h_{r,s}^{P}={rs\over m}+{c\over 24},~~~q_{r,s}^P=(r-s)/m.}

In the T sector the characters are
\eqn\twist{\chi^{(N=2,T)(m)}_r(q)=
{(-q^{1\over2})_{\infty}(-q)_{\infty}\over (q^{1\over
2})_{\infty}(q)_{\infty}}\sum_{j=-\infty}^{\infty}\biggl(q^{mj^2+{j\over
2}(m-2r)}-q^{(jm-r)(j-{1\over 2})}\biggr)}
where $1\leq r\leq m/2$
and
\eqn\ntdimt{h_r^{T}={(m-2r)^2\over 16m}+{c\over 24}.}

In the $A$ and $P$ sectors these characters (with $m=3$) are all
obtained from the Bailey pairs A(5) and A(8) of Slater by specializing
{}~\isingc~to the point
\eqn\specnt{aq/\rho_1\rho_2=1.}
To carry out this specialization rewrite ~\isingc~using
\eqn\summand{\eqalign{&{(\rho_1)_{3j}(\rho_2)_{3j}
(aq/\rho_1\rho_2)^{3j}\over
(aq/\rho_1)_{3j}(aq/\rho_2)_{3j}}-
{(\rho_1)_{3j+1}(\rho_2)_{3j+1}(aq/\rho_1\rho_2)^{3j+1}\over
(aq/\rho_1)_{3j+1}(aq/\rho_2)_{3j+1}}\cr
&={(\rho_1)_{3j}(\rho_2)_{3j}
(aq/\rho_1\rho_2)^{3j}\over
(aq/\rho_1)_{3j}(aq/\rho_2)_{3j}}
\biggl(1-{(1-\rho_1q^{3j})(1-\rho_2q^{3j})(aq/\rho_1\rho_2)\over
(1-aq^{3j+1}/\rho_1)(1-aq^{3j+1}/\rho_2)}\biggr)\cr
=&{(\rho_1)_{3j}(\rho_2)_{3j}
(aq/\rho_1\rho_2)^{3j}\over
(aq/\rho_1)_{3j}(aq/\rho_2)_{3j}}{(1-aq/\rho_1\rho_2)(1-aq^{6j+1})\over
(1-aq^{3j+1}/\rho_1)(1-aq^{3j+1}/\rho_2)}\cr}}
and set
\eqn\specntb{\rho_1=-y^{-1}q^r,~~\rho_2=-yq^s,~~a=q^{r+s-1}}
to obtain
\eqn\specntc{\eqalign{\sum_{n=0}^{\infty}&(-y^{-1}q^r)_n(-yq^s)_n
{q^{n^2}q^{n(r+s-1)}\over
(q^{r+s})_{2n}}\cr
&={(-y^{-1}q^r)_{\infty}(-yq^s)_{\infty}\over
(q^{r+s})_{\infty}(q)_{\infty}}
\sum_{j=-\infty}^{\infty}q^{3j^2+(r+s)j}\biggl(1-{y^{-1}q^{3j+r}\over
1+y^{-1}q^{3j+r}}-{yq^{3j+s}\over 1+yq^{3j+s}}\biggr).\cr}}
The right hand side of this agrees with the character
formula~\ntwo~after multiplying by a suitable factor. Thus in the
$A$ sector we obtain the $q$ series representations
\eqn\nta{\eqalign{\sum_{n=0}^{\infty}&(-y^{-1}q^{1\over 2})_n(-yq^{1\over
2})_n{q^{n^2}\over (q)_{2n}}=\chi_{{1\over 2},{1\over
2}}^{(N=2)(3)}(y,q)\cr
\sum_{n=0}^{\infty}&(-y^{-1}q^{1\over 2})_n(-yq^{1\over 2})_{n+1}{q^{n^2+n}
\over
(q)_{2n+1}}=\chi_{{1\over 2},{3\over 2}}^{(N=2),(3)}(y,q),\cr}}
and in the $P$ sector we find the three characters
\eqn\ntb{\eqalign{\sum_{n=0}^{\infty}&(-y^{-1}q)_n(-yq)_n{q^{n^2+n}\over
(q)_{2n+1}}=\chi_{1,1}^{(N=2)(3)}(y,q)\cr
\sum_{n=0}^{\infty}&(-y^{-1}q)_n(-yq)_{n-1}{q^{n^2}\over
(q)_{2n}}=\chi_{1,0}^{(N=2)(3)}(y,q)\cr
\sum_{n=0}^{\infty}&(-y^{-1}q)_{n+1}(-yq)_{n-1}{q^{n^2+n}\over
(q)_{2n+1}}=\chi_{2,0}^{(N=2)(3)}(y,q)\cr}}

For the $T$ sector we specialize~\isingc~using
\eqn\sptsec{\rho_1=-q^{1\over 2},~~\rho_2=-q,~~a=1}
to find
\eqn\sptsecb{\sum_{n=0}^{\infty}(-q^{1\over 2})_n(-q)_n{q^{n^2-{n\over
2}}\over (q)_{2n}}=\chi_1^{(N=2,T)(3)}(q).}

In a like fashion $q$ series for linear combinations of chacters are
obtained from the Bailey pairs  A(6) and A(7) by
specializing~\isingd. Thus for $\rho_1=-yq^{1\over 2}$
$\rho_2=-y^{-1}q^{1\over 2}$ and $a=1$ we find in the A sector
\eqn\ntnewid{\sum_{n=0}^{\infty}(-y^{-1}q^{1\over 2})_n(-yq^{1\over
2})_n{q^{n(n-1)}\over (q)_{2n}}=\chi_{{1\over 2}, {3\over 2}}^{(N=2)(3)}(y,q)
+\chi_{{1\over 2}, {3\over 2}}^{(N=2)(3)}(y^{-1},q),}
and for $\rho_1=-y^{-1}q,$ $\rho_2=-yq$ and $a=q$ we find in the P sector
\eqn\ntnewidb{\sum_{n=0}^{\infty}(-y^{-1}q)_n(-yq)_n{q^{n^2}\over
(q)_{2n+1}}={1\over
2}\{\chi_{1,0}^{(N=2)(3)}(y,q)+\chi_{1,0}^{(N=2)(3)}(y^{-1},q)\}.}

All of these fermionic $q-$series may be put in the canonical
quasi-particle form~\rmccoy~ by use of~\isingl. Thus, for example
we find from~\nta~that
\eqn\ntcan{\eqalign{\sum_{m1,m2,n=0}^{\infty}&y^{m_1-m_2}
q^{{1\over2}(n-m_1)(n-m_1)+{1\over 2}(n-m_2)(n-m_2)+n^2}
{1\over (q)_{2n}}{n\atopwithdelims[] m_1}_q
{n\atopwithdelims[] m_2}_q\cr
&=\chi_{{1\over 2},{1\over 2}}^{(N=2)(3)}(y,q)\cr}}
which if we use $n=m_3/2$ is rewritten as
\eqn\dform{\sum_{{\vec m}=0,m_3~{\rm even}}^{\infty}
q^{{1\over 4}{\vec m}C_{D_3}{\vec m}}y^{m_1-m_2}{1\over
(q)_{m_3}}{{1\over 2}m_3\atopwithdelims[] m_1}_q{{1\over
2}m_3\atopwithdelims[] m_2}_q=\chi_{{1\over 2},{1\over
2}}^{(N=2)(3)}(y,q)}
where $C_{D_3}$ is the Cartan matrix of $D_3.$
This form of the $N=2$ characters was  first given in ~\rmccoy.

\newsec{Supersymmetry from the $F$ pairs of Slater}

Slater~\rslata, in fact, provides not one but two constructions of
the characters of the Ising model. Her second construction is from the
pairs she calls $F$ and
 is quite different from the one presented above, not only in that the
bosonic form of the $M(3,4)$ character is not of the Rocha-Caridi
form~\rocca~ but the specialization~\specb~ which for the $A$ pairs gave $N=1$
supersymmetry now gives $N=2.$

\subsec{The Bailey pairs}

We will  derive a set of Bailey pairs which are somewhat more
general than Slater~\rslata~in that our form will be valid for all
values of $a$ wheras hers are valid only for $a=0$ and $1.$
In particular we start with her equation (2.1)
and set
$e=a,$
 $b=q^{-j},$ and use the identity
\eqn\slatb{{(x)_{\infty}\over (xq^j)_{\infty}}=(x)_j}
to find
\eqn\slata{\sum_{n=0}^{j}{(1-aq^{2n})(q^{-j})_n(c)_n(d)_n(a)_n\over
(1-a)(aq^{1+j})_n(aq/c)_n(aq/d)_n(q)_n}\biggl({aq^{1+j}\over
cd}\biggr)^n={(aq)_j(aq/cd)_j\over(aq/c)_j (aq/d)_j}}
and then use
\eqn\identa{(aq)_{N+n}=(aq)_N(aq^{N+1})_n}
and
\eqn\identb{(xq^{-N})_n=(-1)^nx^nq^{-nN+{1\over 2}n(n-1)}{(q/x)_N\over
(q/x)_{N-n}}}
to obtain
\eqn\slatc{\sum_{n=0}^{j}{(1-aq^{2n})(-1)^nq^{{1\over
2}n(n+1)}a^n(c)_n(d)_n(a)_n\over (1-a) (aq)_{n+j}(q)_{j-n}(q)_n c^n(aq/c)_n
d^n(aq/d)_n}={(aq/cd)_j\over (q)_j (aq/c)_j (aq/d)_j}.}

 There are many specializations of this result which will give Bailey
pairs. For example if $c,d\rightarrow \infty$ we obtain Bailey pairs
for the original Rogers-Ramanujan identities for the $M(2,5)$ model.

We here consider first $c=q^{1\over 2}$ and $d\rightarrow \infty$
to find
\eqn\slatd{\sum_{n=0}^{j}{(1-aq^{2n})q^{n^2-{n\over 2}}
a^n(a)_n (q^{1\over 2})_n\over
(1-a)(aq)_{n+j}(q)_{j-n}(q)_n(aq^{1\over 2})_n}=
{1\over (q)_j(aq^{1\over 2})_j}}
and thus comparing with the definition of Bailey pair~\bail~
we find the Bailey pair
\eqn\slate{\eqalign{\alpha_n&={(1-aq^{2n})q^{n^2-{n\over 2}}a^n(a)_n(q^{1\over
2})_n\over (1-a)(q)_n(aq^{1\over 2})_n}\cr
&\beta_n={1\over (q)_n(aq^{1\over2})_n}.\cr}}
We note in particular that Slater gives the  special case $F(1)$ of $a=1$
where
\eqn\slatf{\beta_n={1\over (q^{1\over 2})_n(q)_n},~~
\alpha_n=\cases{1~&for $n=0$\cr
q^{n^2}(q^{n\over 2}+q^{-{n\over 2}})&for $n\geq 1$\cr}}
and $F(2)$ of $a=q$ where
\eqn\slatg{\beta_n={1\over (q^{3\over 2})_n(q)_n},~~~
\alpha_n=q^{n^2+{n\over 2}}{1+q^{n+{1\over 2}}\over
1+q^{1\over 2}}.}

We now use the Bailey~\slate~ pair in Bailey's lemma~\inflemma~and obtain
 \eqn\slath{\eqalign{\sum_{n=0}^{\infty}&
{(\rho_1)_n(\rho_2)_n(aq/\rho_1\rho_2)^n
\over (q)_n(aq^{1\over 2})_n}\cr
&={(aq/\rho_1)_{\infty}(aq/\rho_2)_{\infty}\over
(aq)_{\infty}(aq/\rho_1\rho_2)_{\infty}}\sum_{n=0}^{\infty}
{(\rho_1)_n(\rho_2)_n(aq/\rho_1\rho_2)^n\over (aq/\rho_1)_n(aq/\rho_2)_n}
{(1-aq^{2n})q^{n^2-{n\over 2}}a^n(a)_n(q^{1\over
2})_n\over (1-a)(q)_n(aq^{1\over 2})_n}.\cr}}
As in sec. 2 we will obtain characters of conformal field theory by
considering different specializations of the parameters $\rho_1$, and
$\rho_2.$

\subsec{The case $\rho_1$ and $\rho_2\rightarrow \infty;$ the model $M(3,4)$}

The first specialization to consider is
$\rho_1,\rho_2\rightarrow\infty$ where we find from ~\slath~that
\eqn\slati{\sum_{n=0}^{\infty}{q^{n^2}a^n\over (q)_n(aq^{1\over
2})_n}=
{1\over (aq)_{\infty}}\sum_{n=0}^{\infty}{q^{2n^2-{n\over
2}}a^{2n}(1-aq^{2n})(a)_n(q^{1\over 2})_n\over (1-a)(q)_n(aq^{1\over 2})_n}.}
This result is particularly transparent when $a=1$ where we find
\eqn\slatj{\sum_{n=0}^{\infty}{q^{n^2}\over (q)_n(q^{1\over
2})_n}={1\over
(q)_{\infty}}\sum_{n=-\infty}^{\infty}q^{2n^2}q^{n\over2}}
from which if we finally send $q\rightarrow q^2$ and use the identity
\eqn\slatk{(q^2;q^2)_n(q;q^2)_n=(q)_{2n}}
we obtain
\eqn\slatl{\sum_{n=0}^{\infty}{q^{2n^2}\over
(q)_{2n}}={1\over (q^2;q^2)_{\infty}}
\sum_{n=-\infty}^{\infty}q^{4n^2+n}={1\over (q)_{\infty}(-q)_{\infty}}
\sum_{n=-\infty}^{\infty}q^{4n^2+n}}
which is (39) of Slater~\slatb.
{}From the fermionic side of ~\isinga~we identify this as the character
of the $M(3,4)$ model $\chi_{1,1}^{(3,4)}(q)$ and thus we have
produced another bosonic form for the character which is different
from that of Rocha-Caridi~\rocca~in that the series has no minus signs.
Similarly if $a=q$ we find
\eqn\slatm{\sum_{n=0}^{\infty}{q^{n^2+n}\over (q)_n(q^{3\over 2})_n}
={1\over (q^2)_{\infty}}\sum_{n=0}^{\infty}q^{2n^2+{3\over
2}n}{1+q^{n+{1\over 2}}\over 1+q^{1\over 2}}}
and if we again set $q\rightarrow q^2$ and use
\eqn\slatn{(q^2;q^2)_n(q^3;q^2)_n={(q)_{2n+1}\over 1-q}}
we find
\eqn\slato{\sum_{n=0}^{\infty}{q^{2n(n+1)}\over
(q)_{2n+1}}={1\over
(q^2;q^2)_{\infty}}\sum_{n=0}^{\infty}q^{4n^2+3n}(1+q^{2n+1})={1\over
(q^2;q^2)_{\infty}}\sum_{n=-{\infty}}^{\infty}q^{4n^2+3n}}
where the final expression is obtained from the previous sum by using
$n\rightarrow -n-1$ in the second term.
This identity is (38) of Slater~\rslatb~and on comparison of the fermionic
side  with ~\isinga~is seen to be the character $\chi_{1,3}^{(3,4)}(q).$

We have thus obtained new bosonic sum expressions for the characters
$\chi_{1,1}^{(3,4)}(q)$ and $\chi_{1,3}^{(3,4)}(q)$ of the model
$M(3,4).$ It remains to find the bosonic forms for the character
$\chi_{1,2}^{(3,4)}(q).$ This is done~\rslata~by returning to
{}~\slatc~and setting $c=q^{1\over 2}$ and $d=0$ to find
\eqn\slatp{\sum_{n=0}^{j}{(1-aq^{2n})q^{-{n\over 2}}(q^{1\over
2})_n(a)_n\over (1-a)(aq)_{n+j}(q)_{j-n}(q)_n(aq^{1\over
2})_n}={q^{-{j\over 2}}\over (q)_j(aq^{1\over 2})_j}.}
Thus we find the Bailey pairs
\eqn\slatq{\alpha_n={(1-aq^{2n})q^{-{n\over 2}} (q^{1\over
2})_n(a)_n\over (1-a)(q)_n(aq^{1\over 2})_n},~~~~\beta_n={q^{-{n\over 2}}\over
(q)_n(aq^{1\over 2})_n}}
and in particular note that for $a=1$ we have
\eqn\slatr{\beta_n={q^{-{n\over 2}}\over (q)_n(q^{1\over2})_n},
{}~~~~\alpha_n=\cases{1&for $n=0$\cr
q^{n\over 2}+q^{-{n\over 2}}&~for $ n\geq 1$}}
which is F(3) of Slater~\rslata~and setting $a=q$ we find
\eqn\slats{\beta_n={q^{-{n\over 2}}\over (q)_n(q^{3\over 2})_n},~~~~~
\alpha_n=q^{-{n\over 2}}{1+q^{n+{1\over 2}}\over 1+q^{1\over 2}}}
which is F(4) of Slater. Thus if we use this in Bailey's lemma with
$N\rightarrow \infty$ we find
\eqn\slatt{\eqalign{\sum_{n=0}^{\infty}&{(\rho_1)_n(\rho_2)_n
(aq^{1\over 2}/\rho_1\rho_2)^n\over
(q)_n(aq^{1\over 2})_n}\cr
&={(aq/\rho_1)_{\infty}(aq/\rho_2)_{\infty}\over
(aq)_{\infty}(aq/\rho_1\rho_2)_{\infty}}\sum_{n=0}^{\infty}
{(\rho_1)_n(\rho_2)_n(aq/\rho_1\rho_2)^n\over
(aq/\rho_1)_n(aq/\rho_2)_n}{(1-aq^{2n})q^{-{n\over 2}}
(q^{1\over 2})_n(a)_n\over (1-a)(q)_n(aq^{1\over2 })_n}\cr}}
and then sending $\rho_1,\rho_2\rightarrow \infty$ we find
\eqn\slatu{\sum_{n=0}^{\infty}{q^{n^2-n/2}a^n\over (q)_n(aq^{1\over
2})_n}={1\over (aq)_{\infty}}\sum_{n=0}^{\infty}{q^{n^2-{n\over
2}}a^{n}(1-aq^{2n})(a)_n(q^{1\over 2})_n\over (1-a)(q)_n
(aq^{1\over 2})_n}.}
Then if $a=1$ and $q\rightarrow q^2$ we use ~\slath~ to find
\eqn\slatv{\sum_{n=0}^{\infty}{q^{n(2n-1)}\over (q)_{2n}}=
{1\over
(q^2;q^2)_{\infty}}\sum_{n=-\infty}^{\infty}q^{2n^2}q^n}
where from~\isinga~we identify the fermionic side as
$\chi_{1,2}^{(3,4}(q).$
Similarly if $a=q$ and $q\rightarrow q^2$ we find
\eqn\slatw{\sum_{n=0}^{\infty}{q^{n(2n+1)}\over (q)_{2n+1}}={1\over
(q^2;q^2)_{\infty}}\sum_{n=0}^{\infty}q^{2n^2+n}(1+q^{2n+1})}
and we note that if in the second term of the bosonic sum we let
$n\rightarrow -n-1$ we regain the bosonic sum of~\slatv. This agrees
with the equality of the characters
$\chi_{1,2}^{(3,4)}(q)$  and $\chi_{2,2}^{(3,4)}(q)$ which is seen in
{}~\isinga~and also agrees with (9) of Slater~\rslatb.

\subsec{The case $\rho_1\rightarrow \infty$ and  $\rho_2$ finite
;~$N=2$ supersymmetry}

We next let $\rho_1\rightarrow\infty$ and $\rho_2=-q^{1\over 2}.$
Then
for the Bailey pair~\slate~we obtain from ~\slath
\eqn\slatx{\sum_{n=0}^{\infty}{q^{n^2\over 2}a^n(-q^{1\over 2})_n\over
(q)_n(aq^{1\over 2})_n}={(-aq^{1\over 2})_{\infty}\over
(aq)_{\infty}}\sum_{n=0}^{\infty}{q^{{3n^2\over 2}-{n\over
2}}a^{2n}(1-aq^{2n})(a)_n(q^{1\over 2})_n(-q^{1\over
2})_n\over (1-a)(q)_n(aq^{1\over 2})_n(-aq^{1\over 2})_n}.}
Then if $a=1$ and $q\rightarrow q^2$
we find
\eqn\slaty{\sum_{n=0}^{\infty}{q^{n^2}(-q;q^2)_n\over
(q)_{2n}}={(-q;q^2)_{\infty}\over (q^2;q^2)_{\infty}}
\sum_{n=-\infty}^{\infty}q^{{3n^2}+n}}
where the fermionic side is recognized as (29) of
Slater~\rslatb. Moreover, if we use the identity
\eqn\slatid{\eqalign{(-q;q^2)_n&=\prod_{l=0}^{n-1}(1+q^{2l+1})=
\prod_{n=0}^{n-1}
(1+iq^{1\over 2}q^l)(1-iq^{1\over 2}q^l)\cr
&=(iq^{1\over 2})_n(-iq^{1\over 2})_n\cr}}
we may rewrite the fermionic side to obtain
\eqn\slatida{\sum_{n=0}^{\infty}q^{n^2}{(iq^{1\over 2})_n
(-iq^{1\over 2})_n\over(q)_{2n}}= {(-q;q^2)_{\infty}\over (q^2;q^2)_{\infty}}
\sum_{n=-\infty}^{\infty}q^{{3n^2}+n}}
and thus comparing the fermionic side with ~\nta~ with $y=i$ we
identify this with the $N=2$ supersymmetric character $\chi_{{1\over
2},{1\over 2}}^{(N=2) (m=3)}(i,q).$ This identification has not
appeared in the literature before.

Similarly if
$a=q$ and $q\rightarrow q^2$ we find the $q-$series identity
\eqn\slatz{\sum_{n=0}^{\infty}{q^{n^2+2n}(-q;q^2)_n\over
(q)_{2n+1}}=\sum_{n=0}^{\infty}q^{n^2+2n}{(iq^{1\over 2})_n (-iq^{1\over
2})_n\over(q)_{2n+1}}
={(-q;q^2)_{\infty}\over
(q^2;q^2)_{\infty}}\sum_{n=0}^{\infty}q^{3n^2+3n}.}
This last expression, however, is not a theta function and lacks the
physical interpretation as a character.

We may also consider the Bailey pair
{}~\slate~with $\rho_1 \rightarrow \infty$ and  $\rho_2=-q$
where we find from ~\slath
\eqn\slataa{\sum_{n=0}^{\infty}{q^{{1\over 2}n(n-1)}a^n(-q)_n\over
(q)_n(aq^{1\over 2})_n}={(-a)_{\infty}\over
(aq)_{\infty}}\sum_{n=0}^{\infty}{q^{{3n^2\over
2}-n}a^{2n}(1-aq^{2n})(a)_n(q^{1\over 2})_n(-q)_n\over
(1-a)(q)_n(aq^{1\over 2})_n(-a)_n}.}
Then setting $a=1$ and $q\rightarrow q^2$ we get
\eqn\slatab{\sum_{n=0}^{\infty}{q^{n(n-1)}(-q^2;q^2)_n\over
(q)_{2n}}=\sum_{n=0}^{\infty}q^{n(n-1)}{(iq)_n(-iq)_n\over (q)_{2n}}
={(-q^2;q^2)_{\infty}\over (q^2;q^2)_{\infty}}
\sum_{n=0}^{\infty}q^{3n^2}(q^{n}+q^{-n})^2}
which is also an identity is not found in Slater and  which lacks a character
interpretation.
Similarly if $a=q$ and $q\rightarrow q^2$ we find
\eqn\slatac{\eqalign{\sum_{n=0}^{\infty}{q^{n(n+1)}(-q^2;q^2)_n\over
(q)_{2n+1}}&=\sum_{n=0}^{\infty}{q^{n(n+1)}(iq)_n(-iq)_n\over
(q)_{2n+1}}={(-q^2;q^2)_{\infty}\over(q^2;q^2)_{\infty}}
\sum_{n=0}^{\infty}q^{3n^2+2n}(1+q^{2n+1})\cr
&={(-q^2;q^2)_{\infty}\over (q^2;q^2)_{\infty}}
\sum_{n=-\infty}^{\infty}q^{3n^2+2n}\cr}}
Here the fermionic side is (28) of Slater and is also the
fermionic side of $\chi_{1,1}^{(N=2)(3)}(i,q)$ ~\ntb.

Similarly for the Bailey pair ~\slatq~ if we set
$\rho_1\rightarrow \infty$ and  $\rho_2=-q^{1\over 2}$
in ~\slatt~we find
\eqn\slatad{\sum_{n=0}^{\infty}{q^{{n\over 2}(n-1)}a^n(-q^{1\over
2})_n\over (q)_n(aq^{1\over 2})_n}=
{(-aq^{1\over 2})_{\infty}\over(aq)_{\infty}}\sum_{n=0}^{\infty}
{q^{{n^2\over 2}-{n\over
2}}a^{n}(1-aq^{2n})(a)_n(q^{1\over 2})_n(-q^{1\over
2})_n\over (1-a)(q)_n(aq^{1\over 2})_n(-aq^{1\over 2})_n.}}
Thus, if $a=1$ and $q\rightarrow q^2$ we find
\eqn\slatae{\sum_{n=0}^{\infty}{{q^{n(n-1)}(-q;q^2)_n\over
(q)_{2n}}}=\sum_{n=0}^{\infty}q^{n(n-1)}{(iq^{1\over2})_n
(-iq^{1\over 2})_n\over (q)_{2n}}=
{(-q;q^2)_{\infty} \over
(q^2;q^2)_{\infty}}\sum_{n=-\infty}^{\infty}q^{n^2+n}}
and by comparison of the fermionic side with ~\ntnewid~ we see that this
is the sum of characters $\chi_{{1\over 2},{3\over
2}}^{(N=2)(3)}(i,q)+\chi_{{3\over 2},{1\over 2}}^{(N=2)(3)}(i,q).$
This result is not found in Slater ~\rslatb.

Finally we set $\rho_1=-q$ and $\rho_2\rightarrow \infty$ in
{}~\slatt~and
find for $a=q$ and $q\rightarrow q^2$
\eqn\slatah{\sum_{n=0}^{\infty}{q^{n^2}(-q^2;q^2)_{n}\over (q)_{2n+1}}=
\sum_{n=0}^{\infty}q^{n^2}{(iq)_n(-iq)_n\over (q)_{2n+1}}=
{(-q^2;q^2)_{\infty}\over (q^2;q^2)_{\infty}}\sum_{n=-\infty}^{\infty}q^{n^2}}
where the fermionic side is identical with that of ~\ntnewidb~with $y=i$
and thus is equal to
${1\over 2}\biggl(\chi_{1,0}^{(N=2)(3)}(i,q)+
\chi_{1,0}^{(N+2)(3)}(-i,q)\biggr).$
This result also is not found in Slater.



\newsec{$N=2$ supersymmetric models from $M(p-1,p)$}

The derivation given in the previous section of the characters of a
$N=1$ and $N=2$ supersymmetric model from the $M(3,4)$ minimal model
by means of the Bailey lemma may be generally applied to the $M(p,p')$
model. We will however restrict ourselves here to the unitary case
$M(p-1,p)$ and treat the general case in~\rbms.
In order to somewhat simplify our treatment we will first extend
Bailey's lemma~\lemma~to what we will call the bilateral Bailey lemma.
We will then demonstrate how from the Bailey pair obtained from the
finite size character polynomials for $M(p-1,p)$ we may construct the
characters for $M(p,p+1).$ This is an example of the interpretation of
renormalization flows~\razam--\rfqr~in terms of Bailey pairs mentioned in
{}~\rbm. We then generalize the method of sec. 2 to obtain
the $N=1$ and $N=2$ supersymmetric models by the
specialization of parameters $\rho_1$ and $\rho_2.$

\subsec{Bilateral Bailey Pairs}

The pair $(\alpha_n,\beta_n)$ is said to be a bilateral Bailey pair if
the following relation is satisfied
\eqn\bibail{\beta_n=\sum_{j=-\infty}^{n}{\alpha_j\over (q)_{n-j}(aq)_{n+j}}}
where $(a)_{-n}$ is defined as in~\min.
One can easily check that with this definition the following important
properties of $(a)_n$ hold for positive and negative integer $n,k$
\eqn\prop{\eqalign{{(a)_n\over (a)_{n-k}}&=(-1)^k(a/q)^kq^{nk-{1\over 2}k(k-1)}
(q^{1-n}/a)_k,\cr
(a)_{n+k}&=(a)_n(aq^{n})_k.\cr}}
Analogous to the proof of the original Bailey lemma~\lemma~ given in~\rbail~
and~\rand~one may derive the
bilateral Bailey lemma using~\prop~which states that if $(\alpha_n,\beta_n)$
satisfy~\bibail~then
\eqn\bilemma{\eqalign{\sum_{n=-\infty}^{N}&\biggl({(\rho_1)_n
(\rho_2)_n(aq/\rho_1
\rho_2)^n \alpha_n\over (aq/\rho_1)_n (aq/\rho_2)_n}\biggr)
{1\over (q)_{N-n}(aq)_{N+n}}\cr
&~~~=\sum_{n=-\infty}^{N}{(\rho_1)_n (\rho_2)_n(aq/\rho_1
\rho_2)_{N-n}(aq/\rho_1 \rho_2)^n\beta_n\over (q)_{N-n}(aq/\rho_1)_N
(aq/\rho_2)_N}}}
given that the series converge. In analogy to~\inflemma~ we let
$N\rightarrow\infty$ to obtain
\eqn\infbilemma {\sum_{n=-\infty}^{\infty}
(\rho_1)_n(\rho_2)_n(aq/\rho_1\rho_2)^n\beta_n\
={(aq/\rho_1)_{\infty}(aq/\rho_2)_{\infty}\over
(aq)_{\infty}(aq/\rho_1\rho_2)_{\infty}}\sum_{n=-\infty}^{\infty}\biggl(
{(\rho_1)_n (\rho_2)_n(aq/\rho_1\rho_2)^n \alpha_n\over (aq/\rho_1)_n
(aq/\rho_2)_n}\biggr).}

Following Andrews~\rand~one may introduce dual (bilateral) Bailey pairs. If
$(\alpha_n,\beta_n)$ is a (bilateral) Bailey pair relative to $a$ then
\eqn\dual{\eqalign{A_n(a,q)&=a^nq^{n^2}\alpha_n(a^{-1},q^{-1}),\cr
B_n(a,q)&=a^{-n}q^{-n^2-n}\beta_n(a^{-1},q^{-1})\cr}}
also satisfy~\bail~(\bibail) relative to $a$ and $(A_n,B_n)$ is called
the dual (bilateral) Bailey pair to $(\alpha_n,\beta_n)$.

Foda-Quano~\rfqa~ used the polynomial Fermi-Bose character identity for the
$M(p-1,p)$ model to derive Bailey pairs. We will quickly review their
method and slightly generalize it to obtain Bailey pairs that yield the
characters for the $M(p,p+1)$ model, the $SM(p,p+2)$ model and the $N=2$
supersymmetric model with central charge $c=3(1-2/p)$.

The Bose-Fermi
character polynomial identities for the minimal models $M(p-1,p)$ are of
the form
\eqn\febo{B_{r,s}^{(L,p)}=F_{r,s}^{(L,p)}}
where $B_{r,s}^{(L,p)}$ is the function of Andrews, Baxter and Forrester
{}~\rabf
\eqn\bose{\eqalign{B_{r,s}^{(L,p)}=\sum_{j=-\infty}^{\infty}&\biggl(
 q^{j(jp(p-1)+pr-(p-1)s)}{L\atopwithdelims[] [{1\over 2}(L+s-r)]-pj}_q\biggr.
\cr
 &\biggl.-q^{(jp-s)(j(p-1)-r)}{L\atopwithdelims[] [{1\over 2}(L-s-r)]+pj}_q
 \biggr). }}
Here ${n\atopwithdelims[] j}_q$ are the q-binomial coefficients defined in
{}~\qbin~and $[x]$ denotes the integer part of $x$.
Equation~\bose~can be put in the form of~\bibail~by setting $L=2l+r-s+2x$
in~\bose~ and using
\eqn\deriv{\eqalign{{2l+r-s+2x\atopwithdelims[] l+x-pj}_q&=
{(q)_{2l+r-s+2x}\over (q)_{l-(pj-x)}(q)_{l+r-s+x+pj}}\cr
&={(q^{r-s+2x+1})_{2l}\over (q)_{l-(pj-x)}(q^{r-s+2x+1})_{l+(pj-x)}},\cr
&\cr
{2l+r-s+2x\atopwithdelims[] l+x-s+pj}_q&=
{(q^{r-s+2x+1})_{2l}\over (q)_{l-(pj-r-x)}(q^{r-s+2x+1})_{l+(pj-r-x)
}}.\cr}}
Thus we can read off the Bailey pair relative to $a=q^{r-s+2x}$ from~\febo~as
\eqn\bpe{\eqalign{\alpha_n&=\cases{q^{j(jp(p-1)+pr-(p-1)s)} & for $n=pj-x$\cr
 -q^{(jp-s)(j(p-1)-r)} & for $n=pj-r-x$\cr 0& otherwise\cr}\cr&\cr
\beta_n&=\cases{{1\over (aq)_{2n}}
 F_{r,s}^{(2n+r-s+2x,p)}(q)& for $n\geq 0$\cr 0&otherwise.\cr}\cr}}

Similarly if we set $L=2l+r-s+2x+1$ we get the Bailey pair relative to
$a=q^{r-s+2x+1}$
\eqn\bpo{\eqalign{\alpha_n&=\cases{q^{j(jp(p-1)+pr-(p-1)s)} & for $n=pj-x$\cr
 -q^{(jp-s)(j(p-1)-r)} & for $n=pj-r-x-1$\cr 0& otherwise\cr}\cr&\cr
\beta_n&=\cases{{1\over (aq)_{2n}}
 F_{r,s}^{(2n+r-s+2x+1,p)}(q)& for $n\geq 0$\cr 0&otherwise.\cr}\cr}}

In addition the dual Bailey pairs are obtained from the Bailey dual
construction~\dual.
Accordingly we get the two  dual Bailey pairs, one relative to
$a=q^{r-s+2x}$ (from~ \bpe)
\eqn\bped{\eqalign{\alpha_n&=\cases{q^{j^2p-sj}
q^{x(s-r-x)} & for $n=pj-x$\cr
 -q^{j^2p-sj}q^{x(s-r-x)} & for $n=pj-r-x$
\cr 0& otherwise\cr}\cr &\cr
\beta_n&=\cases{{1\over (aq)_{2n}}q^{n^2}a^{n}
 F_{r,s}^{(2n+r-s+2x,p)}(q^{-1})& for $n\geq 0$\cr 0&otherwise\cr}
\cr}}
and the second relative to $a=q^{r-s+2x+1}$ (from ~\bpo)
\eqn\bpod{\eqalign{\alpha_n&=\cases{q^{j^2p+pj-js}q^{x(s-r-x-1)} & for
$n=pj-x$\cr
 -q^{(j-1)(jp-s)}q^{x(s-r-x-1)} & for $n=pj-r-x-1$\cr 0& otherwise\cr}\cr &\cr
\beta_n&=\cases{{1\over (aq)_{2n}}q^{n^2}a^{n}
 F_{r,s}^{(2n+r-s+2x+1,p)}(q^{-1})& for $n\geq 0$\cr 0&otherwise.\cr}
\cr}}

Thus far we have explicitly constructed only the bosonic side of the
Bailey pair. The fermionic side for the $M(p-1,p)$ model
\eqn\fermi{\eqalign{&F_{r,s}^{(L,p)}
=q^{-(s-r)(s-r-1)/4}\sum_{{\vec m}\equiv
 {\vec Q}_{r,s}^{(p-3)}}q^{{1\over 4}{\vec m}C_{p-3}{\vec m}
-{1\over 2}{\vec A}_{r,s}^{(p-3)}{\vec m}}
{{1\over 2}(m_2+L+(\vec{u}_{r,s}^{(p-3)})_1)\atopwithdelims[] m_1}_q\cr
&~~~~~\times
\biggl(\prod_{i=2}^{p-4}{{1\over 2}(m_{i-1}+m_{i+1}+(\vec{u}_{r,s}^{(p-3)})_i)
 \atopwithdelims[] m_i}_q\biggr)
{{1\over 2}(m_{p-4}+(\vec{u}_{r,s}^{(p-3)})_{p-3})\atopwithdelims[]
m_{p-3}}_q\cr}}
has been proven in~\rber~,~\rwara-\ranne. Here $C_{p-3}$ is the
$(p-3)\times (p-3)$ dimensional Cartan
matrix of the Lie algebra $A_{p-3}$ with the elements
$C_{j,k}=2\delta_{j,k}-\delta_{j-1,k}-\delta_{j+1,k}$. Furthermore we
define the
vectors ${\vec e}_i$ as the $(p-3)$-dimensional vectors of unit length in the
$i^{\rm th}$ direction
\eqn\unit{({\vec e}_i)_j=\cases{\delta_{i,j} & if $1\leq j\leq p-3$ \cr 0 &
otherwise\cr}}
The vectors ${\vec A}_{r,s}^{(p-3)}, {\vec u}_{r,s}^{(p-3)}\in {\bf N}^{p-3}$
and ${\vec Q}_{r,s}^{(p-3)}\in ({\bf Z}_2)^{p-3}$. The sum in~\fermi~runs
over ${\vec m} \in {\bf N}^{p-3}$ such that ${\vec m}\equiv
{\vec Q}_{r,s}^{(p-3)} \bmod 2$.
When $L+r-s$ even
\eqn\down{\eqalign{{\vec A}_{r,s}^{(p-3)}&={\vec e}_{s-1}\cr
 {\vec u}_{r,s}^{(p-3)}&={\vec e}_{s-1}+{\vec e}_{p-r-1}\cr
 {\vec Q}_{r,s}^{(p-3)}&=(r-1){\vec \rho}^{(p-3)}
 +({\vec e}_{s-2}+{\vec e}_{s-4}+\ldots)
 +({\vec e}_{p-r}+{\vec e}_{p+2-r}+\ldots)\cr}}
and when $L+r-s$ odd
\eqn\up{\eqalign{{\vec A}_{r,s}^{(p-3)}&={\vec e}_{p-s-1}\cr
 {\vec u}_{r,s}^{(p-3)}&={\vec e}_{p-s-1}+{\vec e}_{r}\cr
 {\vec Q}_{r,s}^{(p-3)}&=(s-1){\vec \rho}^{(p-3)}
 +({\vec e}_{r-1}+{\vec e}_{r-3}+\ldots)
 +({\vec e}_{p-s}+{\vec e}_{p+2-s}+\ldots)\cr}}
where ${\vec \rho}^{(p-3)}={\vec e}_1+\ldots+{\vec e}_{p-3}$.

Using
\eqn\limit{\lim_{L\rightarrow\infty}{L\atopwithdelims[] n}={1\over (q)_n}}
one may take the limit $L\rightarrow\infty$ in~\fermi
\eqn\fermilim{F_{r,s}^{(\infty,p)}=q^{-{1\over 4}(s-r)(s-r-1)}
\sum_{\vec{m}\equiv\vec{Q}_{r,s}^{(p-3)}}q^{{1\over 4}\vec{m}C_{p-3}\vec{m}
-{1\over 2}\vec{A}_{r,s}^{(p-3)}\vec{m}}
{1\over (q)_{m_1}}\prod_{i=2}^{p-3}
{{1\over 2}(I_{p-3}\vec{m}+\vec{u}_{r,s}^{(p-3)})_i\atopwithdelims[] m_i}}
where we introduced the incidents matrix $I_{p-3}=2-C_{p-3}$ for compact
notation. Notice that there are two ways to take the limit $L\rightarrow
\infty$, such that $L+r-s$ even or such that $L+r-s$ odd. In the limit
$L+r-s$ even $\vec{A}_{r,s}^{(p-3)}, \vec{u}_{r,s}^{(p-3)},
\vec{Q}_{r,s}^{(p-3)}$ are given as
in~\down~and for $L+r-s$ odd as in~\up. Both limits yield the same $q$-series
and the two different sets for $\vec{A}_{r,s}^{(p-3)}, \vec{u}_{r,s}^{(p-3)}$
and $\vec{Q}_{r,s}^{(p-3)}$ reflect the symmetry of the characters
$\chi_{r,s}^{(p-1,p)}=\chi_{p-1-r,p-s}^{(p-1,p)}$.

We now use the dual bilateral Bailey pairs~\bped~and ~\bpod~
in the bilateral Bailey lemma~\infbilemma~. Thus if we use~\bped~in
the bilateral Bailey
lemma~\infbilemma~we find
\eqn\unilem{\eqalign{&{(aq/\rho_1)_{\infty}(aq/\rho_2)_{\infty}\over
(aq)_{\infty}(aq/\rho_1\rho_2)_{\infty}}\sum_{j=-\infty}^{\infty}
q^{j(jp-s)}q^{x(s-r-x)}\cr
&\times\biggl(
{(\rho_1)_{pj-x}(\rho_2)_{pj-x}(aq/\rho_1\rho_2)^{pj-x}
\over (aq/\rho_1)_{pj-x}
(aq/\rho_2)_{pj-x}}-{(\rho_1)_{pj-r-x}(\rho_2)_{pj-r-x}
(aq/\rho_1\rho_2)^{pj-r-x}\over
(aq/\rho_1)_{pj-r-x} (aq/\rho_2)_{pj-r-x}}\biggr)\cr
&=\sum_{n=0}^{\infty}(\rho_1)_n(\rho_2)_n(aq/\rho_1\rho_2)^n
{q^{n^2}a^n\over (aq)_{2n}}F_{r,s}^{(2n+r-s+2x,p)}(q^{-1})\cr}}
where $a=q^{r-s+2x}$. A very similar equation holds if we insert~\bpod~in
the bilateral Bailey lemma~\infbilemma.

We now need to consider three specializations of the
parameters $\rho_1,\rho_2$ as done in section 2.

\subsec{The model $M(p,p+1)$}

The first specialization  is
\eqn\genspo{\rho_1\rightarrow\infty,~~\rho_2\rightarrow\infty.}
Then if we also set $x=0$ we obtain from~\unilem~with $a=q^{r-s}$
\eqn\unilema{\eqalign{&{1\over (q^{r-s+1})_{\infty}}\sum_{j=-\infty}^{\infty}
\biggl(q^{j(jp(p+1)+rp-s(p+1))}-q^{(jp-s)(j(p+1)-r)}\biggr)\cr
&=\sum_{n=0}^{\infty}{q^{2n^2+2n(r-s)}\over (q^{r-s+1})_{2n}}
F_{r,s}^{(2n+r-s,p)}(q^{-1})}}
and hence, comparing with ~\rocca~we find
\eqn\unilemb{\chi_{r,s}^{(p+1,p)}=\chi_{s,r}^{(p,p+1)}=
\sum_{n=0}^{\infty}{q^{2n(n+r-s)}\over
(q)_{2n+r-s}}F_{r,s}^{(2n+r-s,p)}(q^{-1}).}
Moreover since we started out with
the model $M(p-1,p)$ $r$ and $s$ are restricted to $1\leq r\leq p-2,~
1\leq s\leq p-1$ whereas we need~\unilemb~for $1\leq r\leq p$ and
$1\leq s\leq p-1$. Thus, the range of $r$ needed for $\chi_{s,r}^{(p,p+1)}$
is larger than the range of $r$ for which~\unilemb~holds.

To get the remaining characters we make use
of the Bailey pair~\bpod~with $x=0$ and $a=q^{r-s+1}.$ Then using the
bilateral Bailey lemma~\infbilemma~with $\rho_1,\rho_2\rightarrow\infty$ we
obtain
\eqn\unilemc{\chi_{s,r+2}^{(p,p+1)}=\sum_{n=0}^{\infty}{q^{2n(n+r-s+1)}\over
(q)_{2n+r-s+1}}F_{r,s}^{(2n+r-s+1,p)}(q^{-1}).}

As an application of these results we may check whether~\unilemb~and~\unilemc~
yield the known expressions for the fermionic characters.
To this end let us first compute $F_{r,s}^{(L,p)}(q^{-1})$. Observe that
\eqn\bininv{{n \atopwithdelims[] m}_{q^{-1}}=
q^{m(m-n)}{n \atopwithdelims[] m}_{q}}
from which it follows from~\fermi~that
\eqn\Finv{\eqalign{F_{r,s}^{(L,p)}(q^{-1})=
q^{(s-r)(s-r-1)/4}&\sum_{\vec{m}\equiv \vec{Q}_{r,s}^{(p-3)}}
q^{{1\over 4}\vec{m}C_{p-3}\vec{m}+{1\over 2}\vec{A}_{r,s}^{(p-3)}\vec{m}
-{1\over 2}\vec{u}_{r,s}^{(p-3)}\vec{m}-{1\over 2}Lm_1}\cr
&\prod_{i=1}^{p-3}
{{1\over 2}(I_{p-3}\vec{m}+\vec{u}_{r,s}^{(p-3)}+L\vec{e}_1)_i
\atopwithdelims[] m_i}_q\cr}}
where we introduced the incidence matrix $I_{p-3}=2-C_{p-3}$ for compact
notation.
Hence we obtain from~\unilemb~by defining $m_0=2n+r-s$
\eqn\chiinv{\eqalign{&\chi_{s,r}^{(p,p+1)}=q^{(s-r)(s-r-1)/4}
\sum_{m_0=0,{\rm restr.}}^{\infty}{q^{{1\over 2}m_0^2-{(r-s)^2\over 2}}\over
(q)_{m_0}}\cr
&~~\times\sum_{\vec{m}\equiv \vec{Q}_{r,s}^{(p-3)}}
q^{{1\over 4}\vec{m}C_{p-3}\vec{m}
+{1\over 2}\vec{A}_{r,s}^{(p-3)}\vec{m}-{1\over 2}\vec{u}_{r,s}^{(p-3)}\vec{m}
-{1\over 2}m_0m_1}
\prod_{i=1}^{p-3}{{1\over 2}(I_{p-3}\vec{m}+\vec{u}_{r,s}^{(p-3)}
+m_0\vec{e}_1)_i\atopwithdelims[] m_i}_q\cr}}
where the restrictions on $m_0$ are such that $m_0$ even if $r-s$ even
and $m_0$ odd if $r-s$ odd and $\vec{A}_{r,s}^{(p-3)}, \vec{u}_{r,s}^{(p-3)}$
and $\vec{Q}_{r,s}^{(p-3)}$ as given in
{}~\down. Define $\vec{\tilde{m}}=(m_0,\vec{m})$, $\vec{\tilde{Q}}_{r,s},
\vec{\tilde{u}}_{r,s},\vec{\tilde{A}}_{r,s}$ as in~\down~but
where now
$(\vec{e_i})_j=\delta_{ij}$ for $0\leq i \leq p-3$ and otherwise zero and
$C_{p-2}$ as the $(p-2)$ dimensional Cartan matrix of the Lie algebra
$A_{p-2}$. Accordingly $I_{p-2}=2-C_{p-2}$. Then we have
${1\over 4}\vec{\tilde{m}}C_{p-2}\vec{\tilde{m}}={1\over 2}m_0^2-{1\over 2}
m_0m_1+{1\over 4}\vec{m}C_{p-3}\vec{m}$. Hence
\eqn\chiinva{\chi_{s,r}^{(p,p+1)}=q^{-{1\over 4}(r-s)(r-s-1)}
\sum_{\vec{\tilde{m}}\equiv \vec{\tilde{Q}}_{r,s}}{1\over (q)_{m_0}}
q^{{1\over 4}\vec{\tilde{m}}C_{p-2}\vec{\tilde{m}}+{1\over 2}
\vec{\tilde{A}}_{r,s}
\vec{\tilde{m}}-{1\over 2}\vec{\tilde{u}}_{r,s}\vec{\tilde{m}}}
\prod_{i=1}^{p-3}{{1\over 2}(I_{p-2}\vec{\tilde{m}}+\vec{\tilde{u}}_{r,s})_i
\atopwithdelims[] m_i}_q}
(Notice that we are actually allowed to replace ${1\over 2}
\vec{A}_{r,s}^{(p-3)}\vec{m}-{1\over 2}\vec{u}_{r,s}^{(p-3)}\vec{m}$ in the
exponent by ${1\over 2}\vec{\tilde{A}}_{r,s}\vec{\tilde{m}}
-{1\over 2}\vec{\tilde{u}}_{r,s}\vec{\tilde{m}}$ since $\vec{A}_{r,s}^{(p-3)}
-\vec{u}_{r,s}^{(p-3)}=-\vec{e}_{p-r-1}$ and
$1\leq r\leq p-2$ and hence $\vec{e}_0$ can never be reached).
Defining $\vec{n}=(n_1,n_2,\ldots,n_{p-2})\equiv \vec{\tilde{m}}$ we may
simplify~\chiinva~further to
\eqn\chiinvaa{\chi_{s,r}^{(p,p+1)}=q^{-{1\over 4}(r-s)(r-s-1)}
\sum_{\vec{n}\equiv \vec{Q}_{s,r}^{(p-2)}}{1\over (q)_{n_1}}
q^{{1\over 4}\vec{n}C_{p-2}\vec{n}-{1\over 2}\vec{A}_{s,r}^{(p-2)}\vec{n}}
\prod_{i=2}^{p-2}{{1\over 2}(I_{p-2}\vec{n}+\vec{u}_{s,r}^{(p-2)})_i
\atopwithdelims[] n_i}_q}
where
\eqn\AQU{\eqalign{&\vec{A}_{s,r}^{(p-2)}=\vec{e}_{p-r}\cr
&\vec{u}_{s,r}^{(p-2)}=\vec{e}_s+\vec{e}_{p-r}\cr
&\vec{Q}_{s,r}^{(p-2)}=(r-1)\vec{\rho}^{(p-2)}+(\vec{e}_{s-1}+\vec{e}_{s-3}
+\ldots)+(\vec{e}_{p+1-r}+\vec{e}_{p+3-r}+\ldots).\cr}}
Hence the right hand side of~\chiinvaa~is $F_{s,r}^{(\infty,p+1)}(q)$ with
$\vec{A}_{r,s}^{(p-2)}, \vec{u}_{r,s}^{(p-2)}$ and $\vec{Q}_{r,s}^{(p-2)}$ as
in~\up~(notice that $r$ and $s$ are interchanged in $F_{s,r}^{(\infty,p+1)}$
relative to the definition of $F_{r,s}^{(L,p)}$ in~\fermi).

Similarly one may show that from~\unilemc
\eqn\chiinvb{\chi_{s,r+2}^{(p,p+1)}=F_{s,r+2}^{(\infty,p+1)}(q)}
with $\vec{A}_{r,s}^{(p-2)}, \vec{u}_{r,s}^{(p-2)}$ and
$\vec{Q}_{r,s}^{(p-2)}$ as in~\down.

Thus  all characters of the model $M(p,p+1)$ have been obtained
from the characters of $M(p-1,p)$ by means of the Bailey
construction and hence we have extended the results of ~\rwarb.
This  implementation of the renormalization group
flow  of ~\razam-\rfqr~and is what we call Bailey renormalization flow.

\subsec{The $N=1$ supersymmetric model $SM(p,p+2)$}

The second specialization of~\unilem~is
\eqn\genspob{\rho_1\rightarrow\infty,~~
\rho_2=-q^{r-s+1\over 2}.}
Then if in addition we set $x=0$ so that $a=q^{r-s}$ we see that
$\rho_2=aq/\rho_2$ and all the denominators in~\unilem~cancel. Thus we obtain
from~\unilem~the Neveu-Schwarz characters when $r-s$ is even
\eqn\susy{\hat{\chi}_{r,s}^{(p+2,p)}=\hat\chi_{s,r}^{(p,p+2)}
=\sum_{n=0}^{\infty}{(-q^{1\over 2})_{n+
{r-s\over 2}}\over (q)_{2n+r-s}}q^{{3\over 2}n^2}q^{n{3(r-s)\over 2}}
F_{r,s}^{(2n+r-s,p)}(q^{-1})}
and the Ramond characters when $r-s$ is odd
\eqn\susya{\hat{\chi}_{r,s}^{(p+2,p)}=\hat\chi_{s,r}^{(p,p+2)}
=\sum_{n=0}^{\infty}{(-q)_{n+
{r-s-1\over 2}}\over (q)_{2n+r-s}}q^{{3\over 2}n^2}q^{n{3(r-s)\over 2}}
F_{r,s}^{(2n+r-s,p)}(q^{-1})}
Again it should be mentioned that from the $M(p-1,p)$ construction $r,s$
are restricted by $1\leq r \leq p-2,~1\leq s\leq p-1$ whereas the range
for $r,s$ in $\hat{\chi}_{s,r}^{(p,p+2)}$ is
\eqn\rsrange{1\leq r\leq p+1,~~1\leq s\leq p-1.}

Employing the Bailey pair~\bpod~in the identical fashion with the
specialization
\eqn\newspecial{\rho_1\rightarrow\infty,~~
\rho_2=-q^{r-s+2\over 2},~~
a=q^{r-s+1}}
leads, for the Neveu-Schwarz case when $r-s+3$ even, to
\eqn\susyc{\hat{\chi}_{s,r+3}^{(p,p+2)}=\sum_{n=0}^{\infty}
{(-q^{1\over 2})_{n+{r-s+1\over 2}}\over (q)_{2n+r-s+1}}
q^{{3\over 2}n(n+r-s+1)}F_{r,s}^{(2n+r-s+1,p)}(q^{-1})}
and for the Ramond case, when $r-s+3$ odd, to
\eqn\susyd{\hat{\chi}_{s,r+3}^{(p,p+2)}=\sum_{n=0}^{\infty}
{(-q)_{n+{r-s\over 2}}\over (q)_{2n+r-s+1}}
q^{{3\over 2}n(n+r-s+1)}F_{r,s}^{(2n+r-s+1,p)}(q^{-1}).}
Hence we obtain the fermionic expressions for the whole range in
$r,s$~\rsrange~except for the character $\hat{\chi}_{s,3}^{(4,6)}$.

Again we compare our results~\susy,~\susya,~\susyc,~\susyd~ with
the known results for the fermionic characters for the $SM(p,p+2)$ model as
given in~\rbaver~\ranne~(the fermionic expressions in~\ranne~are actually
expressions for the branching functions and to obtain the characters one
needs to sum over the index $l$)
\eqn\Sfermi{\eqalign{&\hat{\chi}_{r,s}^{(p,p+2)}=\epsilon_{r-s}
q^{-{1\over 8}(s-r-2\epsilon_{r-s}+1)(s-r+2\epsilon_{r-s}-3)}\cr
&~~\times\sum_{m_1\geq 0}\sum_{m_i\equiv (\vec{Q}_{r,s}^{(p-1)})_i,
2\leq i\leq p-1}
q^{{1\over 4}\vec{m}C_{p-1}\vec{m}-{1\over 2}\vec{A}_{r,s}^{(p-1)}\vec{m}}
{1\over (q)_{m_2}}\prod_{i=1,i\not= 2}^{p-1}
{{1\over 2}(I_{p-1}\vec{m}+\vec{u}_{r,s}^{(p-1)})_i\atopwithdelims[] m_i}_q
\cr}}
with $\epsilon_{r-s}$ as in~\epsd, $C_{p-1}$ the $(p-1)$ dimensional Cartan
matrix of $A_{p-1}$, $I_{p-1}=2-C_{p-1}$ and
\eqn\Sdown{\eqalign{&\vec{A}_{r,s}^{(p-1)}=\vec{e}_{s-1}\cr
&\vec{u}_{r,s}^{(p-1)}=\vec{e}_{s-1}+\vec{e}_{p+1-r}
+(2\epsilon_{r-s}-1)\vec{e}_1\cr
&\vec{Q}_{r,s}^{(p-1)}=(r-1)\vec{\rho}^{(p-1)}+(\vec{e}_{s-2}+\vec{e}_{s-4}
+\ldots)+(\vec{e}_{p+2-r}+\vec{e}_{p+4-r}+\ldots)\cr}}
or
\eqn\Sup{\eqalign{&\vec{A}_{r,s}^{(p-1)}=\vec{e}_{p+1-s}\cr
&\vec{u}_{r,s}^{(p-1)}=\vec{e}_{p+1-s}+\vec{e}_{r+1}+(2\epsilon_{r-s}-1)\vec{e}_1\cr
&\vec{Q}_{r,s}^{(p-1)}=(s-1)\vec{\rho}^{(p-1)}+(\vec{e}_{p-s+2}
+\vec{e}_{p-s+4}+\ldots)+(\vec{e}_{r}+\vec{e}_{r-2}+\ldots)\cr}}
where $\vec{\rho}^{(p-1)}=\vec{e}_1+\ldots+\vec{e}_{p-1}$. These two choices
for ${\vec A}_{r,s}^{(p-1)},~{\vec u}_{r,s}^{(p-1)},~{\vec Q}_{r,s}^{(p-1)}$
reflect the symmetry ${\hat \chi}_{r,s}^{(p,p+2)}
={\hat \chi}_{p-r,p+2-s}^{(p,p+2)}.$

To show that~\susy~is of the form~\Sfermi~let us set
$m_0=2n+r-s$ and use~\Finv~and
\eqn\susye{(-q^{1\over 2})_{m_0\over 2}=\sum_{k=0}^{m_0\over 2}
q^{{1\over 2}({m_0\over 2}-k)^2}{{m_0\over 2}\atopwithdelims[] k}_q}
which follows from~\isingl. Then we obtain from~\susy
\eqn\susyf{\eqalign{\hat{\chi}_{s,r}^{p,p+2}&=q^{-{1\over 8}(r-s)(r-s-2)}
\sum_{m_0\geq 0,{\rm even}}^{\infty}\sum_{k=0}^{m_0\over 2}
\sum_{\vec{m}\equiv\vec{Q}_{r,s}^{(p-3)}}q^{{1\over 2}m_0^2-{1\over 2}m_0k
+{1\over 2}k^2-{1\over 2}m_0m_1}\cr
&~~\times q^{{1\over 4}\vec{m}C_{p-3}\vec{m}+{1\over 2}
\vec{A}_{r,s}^{(p-3)}\vec{m}-{1\over 2}\vec{u}_{r,s}^{(p-3)}\vec{m}}
{{m_0\over 2}\atopwithdelims[] k}_q{1\over (q)_{m_0}}_q
\prod_{i=1}^{p-3}{{1\over 2}(I_{p-3}\vec{m}+\vec{u}_{r,s}^{(p-3)}
+m_0\vec{e}_1)_i\atopwithdelims[] m_i}_q\cr}}
where $\vec{A}_{r,s}^{(p-3)}, \vec{u}_{r,s}^{(p-3)}$ and
$\vec{Q}_{r,s}^{(p-3)}$ as given in~\down.
Let us now define $\vec{n}\equiv(n_1,\cdots ,n_{p-1})=
(k,m_0,\vec{m})$, $C_{p-1}$ as the Cartan
matrix of the Lie algebra $A_{p-1}$, and $I_{p-1}=2-C_{p-1}$. Then we may
rewrite~\susyf~as
\eqn\susyg{\eqalign{\hat{\chi}_{s,r}^{p,p+2}&=q^{-{1\over 8}(r-s)(r-s-2)}
\sum_{n_1=0}^{\infty}\sum_{n_i\equiv(\vec{\tilde{Q}}_{r,s}^{(p-1)})_i,
2\leq i\leq p-1}q^{{1\over 4}\vec{n}C_{p-1}\vec{n}+{1\over 2}
\vec{\tilde{A}}_{r,s}^{(p-1)}\vec{n}-{1\over 2}\vec{\tilde{u}}_{r,s}^{(p-1)}
\vec{n}}\cr
&\times{{n_2\over 2}\atopwithdelims[] n_1}_q {1\over (q)_{n_2}}
\prod_{i=3}^{p-1}{{1\over 2}(I_{p-1}\vec{n}+\vec{\tilde{u}}_{r,s}^{(p-1)})_i
\atopwithdelims[] n_i}_q\cr}}
where
\eqn\AuQ{\eqalign{&\vec{\tilde{A}}_{r,s}^{(p-1)}=\vec{e}_{s+1}\cr
&\vec{\tilde{u}}_{r,s}^{(p-1)}=\vec{e}_{s+1}+\vec{e}_{p+1-r}\cr
&\vec{\tilde{Q}}_{r,s}^{(p-1)}=(r-1)\vec{\rho}^{(p-1)}
+(\vec{e}_s+\vec{e}_{s-2}+\ldots)
+(\vec{e}_{p-r+2}+\vec{e}_{p-r+4}+\ldots)\cr}}
with $(\vec{e}_i)_j=\delta_{ij}$ for $1\leq i\leq p-1$ and $\vec{\rho}^{(p-1)}
=\vec{e}_1+\ldots+\vec{e}_{p-1}$. Again, we are allowed to replace
${\vec A}_{r,s}^{(p-3)}{\vec m}-{\vec u}_{r,s}^{(p-3)}{\vec m}$ by
${\vec{\tilde{A}}}_{r,s}^{(p-1)}{\vec n}-\vec{\tilde{u}}_{r,s}^{(p-1)}
{\vec n}$ because
${\vec {\tilde A}}_{r,s}^{(p-1)}-{\vec {\tilde  u}}_{r,s}^{(p-1)}
=-{\vec e}_{p+1-r}$ and
$1\leq r\leq p-2.$
Hence~\susyg~agrees with $\hat{\chi}_{s,r}^{(p,p+2)}$ as in~\Sfermi~with
$\vec{A}_{r,s}^{(p-1)}, \vec{u}_{r,s}^{(p-1)}$ and $\vec{Q}_{r,s}^{(p-1)}$ as
in~\Sup~in the Neveu--Schwarz sector.

Similarly we may treat~\susya~ by using
\eqn\susyh{(-q)_{m_0-1\over 2}={1\over 2}(-1)_{m_0+1\over 2}
={1\over 2}\sum_{k=0}^{m_0+1\over 2}q^{{1\over 2}({m_0+1\over 2}-k)
({m_0-1\over 2}-k)}{{m_0+1\over 2}\atopwithdelims[] k}_q}
and we find the Ramond character
\eqn\chiR{\eqalign{\hat{\chi}_{s,r}^{(p,p+2)}&=
{1\over 2}q^{-{1\over 8}(r-s-1)^2}
\sum_{n_1=0}^{\infty}\sum_{n_i\equiv (\vec{\tilde{Q}}_{r,s}^{(p-1)})_i,
2\leq i\leq p-1}
q^{{1\over 4}\vec{n}C_{p-1}\vec{n}+{1\over 2}\vec{\tilde{A}}_{r,s}^{(p-1)}
\vec{n}-{1\over 2}\vec{\tilde{u}}_{r,s}^{(p-1)}\vec{n}}\cr
&\times{{n_2+1\over 2} \atopwithdelims[] n_1}_q
{1\over (q)_{n_2}}\prod_{i=3}^{p-1}
{{1\over 2}(I_{p-1}\vec{n}+\vec{\tilde{u}}_{r,s}^{(p-1)})_i\atopwithdelims[]
n_i}_q}}
where $\vec{\tilde{A}}_{r,s}^{(p-1)},\vec{\tilde{u}}_{r,s}^{(p-1)},
\vec{\tilde{Q}}_{r,s}^{(p-1)}$ as
in~\AuQ. Hence~\chiR~also agrees with $\hat{\chi}_{s,r}^{(p,p+2)}$ as
in~\Sfermi~with $\vec{A}_{r,s}^{(p-1)},\vec{u}_{r,s}^{(p-1)},
\vec{Q}_{r,s}^{(p-1)}$ as in~\Sup.

Analogously one finds that~\susyc~(\susyd) where
$\vec{A}_{r,s}^{(p-3)},~\vec{u}_{r,s}^{(p-3)},~\vec{Q}_{r,s}^{(p-3)}$ given
by~\up~can be written in the form~\Sfermi~with $\vec{A}_{r,s}^{(p-1)},
\vec{u}_{r,s}^{(p-1)},\vec{Q}_{r,s}^{(p-1)}$ as in~\Sdown.

\subsec{$N=2$ characters with $c=3(1-2/p)$}

Finally we consider the case when
\eqn\choice{\rho_1~{\rm finite},~~~~\rho_2~{\rm finite}.}
Again we distinguish the three sectors $A$, $P$ and $T$.
We start with sector $A$ and set $r=1$ and $\tilde{a}={aq\over \rho_1\rho_2}$
in~\unilem. As in section 2.3 we are interested in the limit $\tilde{a}
\rightarrow 1$. With these definitions the left hand side of~\unilem~becomes
\eqn\ntwoa{\eqalign{&q^{x(s-1-x)}{(\tilde{a}\rho_1)_{\infty}
(\tilde{a}\rho_2)_{\infty}\over
(\tilde{a}\rho_1\rho_2)_{\infty}(\tilde{a})_{\infty}}
\sum_{j=-\infty}^{\infty}{(\rho_1)_{pj-x-1}(\rho_2)_{pj-x-1}\over
(\tilde{a}\rho_1)_{pj-x-1}(\tilde{a}\rho_2)_{pj-x-1}}\tilde{a}^{pj-x-1}
q^{j^2p-sj}\cr
&~~~~~~\times\biggl({(1-\rho_1 q^{pj-x-1})(1-\rho_2 q^{pj-x-1})\over
(1-\tilde{a}\rho_1 q^{pj-x-1})(1-\tilde{a}\rho_2 q^{pj-x-1})}\tilde{a}-1
\biggr)\cr
&=q^{x(s-1-x)}{(\tilde{a}\rho_1)_{\infty}
(\tilde{a}\rho_2)_{\infty}\over
(\tilde{a}\rho_1\rho_2)_{\infty}(\tilde{a})_{\infty}}
\sum_{j=-\infty}^{\infty}
{(\rho_1)_{pj-x-1}(\rho_2)_{pj-x-1}\over
(\tilde{a}\rho_1)_{pj-x-1}(\tilde{a}\rho_2)_{pj-x-1}}\tilde{a}^{pj-x-1}
q^{j^2p-sj}\cr
&~~~~~~~\times
(1-\tilde{a}){\tilde{a}\rho_1\rho_2 q^{2pj-2x-2}-1\over
(1-\tilde{a}\rho_1 q^{pj-x-1})(1-\tilde{a}\rho_2 q^{pj-x-1})}\cr
&\rightarrow_{\tilde{a}\rightarrow 1}q^{x(s-1-x)}
{(\rho_1)_{\infty}(\rho_2)_{\infty}
\over (\rho_1\rho_2)_{\infty}(q)_{\infty}}\sum_{j=-\infty}^{\infty}
q^{j^2p-sj}{\rho_1\rho_2 q^{2pj-2x-2}-1\over
(1-\rho_1 q^{pj-x-1})(1-\rho_2 q^{pj-x-1})}.}}

Let us set
\eqn\ntwob{\rho_1=-yq^{1\over 2},~~~\rho_2=-y^{-1}q^{1\over 2}q^{\rh-\sh}}
where $\rh$ and $\sh$ are half-integers such that $\rh\geq \sh$ and
$s=\rh+\sh$. From this follows that $x=\rh-{1\over 2}$
since ${aq\over \rho_1\rho_2}=1$. Then changing $j\rightarrow -j$ we obtain
from~\ntwoa~and~\unilem
\eqn\ntwoc{\eqalign{&{(-yq^{1\over 2})_{\infty}(-y^{-1}q^{1\over 2})_{\infty}
\over(q)_{\infty}^2}\sum_{j=-\infty}^{\infty}q^{j^2p+(\rh+\sh)j}
{1-q^{2pj+\rh+\sh}\over (1+y^{-1}q^{pj+\rh})(1+yq^{pj+\sh})}\cr
&=q^{-(\rh-{1\over 2})(\sh-{1\over 2})}\sum_{n=0}^{\infty}
{(-yq^{1\over 2})_n(-y^{-1}q^{1\over 2})_{n+\rh-\sh}\over (q)_{2n+\rh-\sh}}
q^{n(n+\rh-\sh)}F_{1,\rh+\sh}^{(2n+\rh-\sh,p)}(q^{-1})}}
which agrees with the $N=2$ unitary characters~\ntwo~
$\chi_{\rh,\sh}^{(N=2)(p)}$ with central charge $c=3(1-2/p)$.

For the $P$ sector we set
\eqn\ntwoP{\rho_1=-yq,~~\rho_2=-y^{-1}qq^{\rh-\sh}}
where now $\rh,\sh$ are integers, $\rh\geq \sh$ and $s=\rh+\sh$. From
${aq\over \rho_1\rho_2}=1$ follows that $x=\rh$. Hence we obtain
\eqn\ntwpPa{\eqalign{&{(-yq)_{\infty}(-y^{-1}q)_{\infty}
\over(q)_{\infty}^2}\sum_{j=-\infty}^{\infty}q^{j^2p+(\rh+\sh)j}
{1-q^{2pj+\rh+\sh}\over (1+y^{-1}q^{pj+\rh})(1+yq^{pj+\sh})}\cr
&=q^{-\rh(\sh-1)}\sum_{n=0}^{\infty}
{(-yq)_n(-y^{-1}q)_{n+\rh-\sh}\over (q)_{2n+\rh-\sh+1}}
q^{n(n+\rh-\sh+1)}F_{1,\rh+\sh}^{(2n+\rh-\sh+1,p)}(q^{-1})}}
which is again in agreement with~\ntwo~in the $P$ sector.

Finally we need to consider the $T$ sector. Here we set $r=1$ and when $s$ odd
\eqn\T{\rho_1=-q^{1\over 2},~~\rho_2=-q,~~a=1.}
Since $a=q^{r-s+2x}$ we read off $x={s-1\over 2}$. Inserting this into the
left hand side of~\unilem~yields
\eqn\Ta{\eqalign{&{(-q^{1\over 2})_{\infty}(-q)_{\infty}\over (q)_{\infty}
(q^{-{1\over 2}})_{\infty}}\sum_{j=-\infty}^{\infty}q^{j(jp-s)}
q^{x(s-1-x)}\cr
&~~~~~~~\times\biggl( (1+q^{pj-x})q^{-{1\over 2}(pj-x)}-(1+q^{pj-x-1})
q^{-{1\over 2}(pj-x-1)}\biggr)\cr
&={(-q^{1\over 2})_{\infty}(-q)_{\infty}\over (q)_{\infty}
(q^{-{1\over 2}})_{\infty}}\sum_{j=-\infty}^{\infty}q^{j(jp-s)}
q^{{1\over 4}(s-1)^2}
\biggl( q^{-{1\over 2}(pj-x)}(1-q^{-{1\over 2}})(-q^{1\over 2}
+q^{pj-x}) \biggr)\cr
&={(-q^{1\over 2})_{\infty}(-q)_{\infty}\over (q)_{\infty}
(q^{1\over 2})_{\infty}}q^{{1\over 4}(s-1)(s-2)}
\sum_{j=-\infty}^{\infty} \biggr( q^{j^2p-js+{1\over 2}pj}
-q^{j^2p-js-{1\over 2}pj+{s\over 2}}\biggl).\cr}}
Hence we obtain
\eqn\Tb{\chi_{s}^{(N=2)(p)}=q^{-{1\over 4}(s-1)(s-2)}
\sum_{n=0}^{\infty}{(-q^{1\over 2})_n(-q)_n\over (q)_{2n}}
q^{n(n-{1\over 2})}F_{1,s}^{(2n,p)}(q^{-1})}
where $\chi_{s}^{(N=2,T)(p)}$ is the character given in~\twist.

For $s$ even we choose
\eqn\Tc{\rho_1=-q^{3\over 2},~\rho_2=-q,~a=q}
which yields $x={s\over 2}$. A similar calculation gives
\eqn\Td{\eqalign{\chi_{s}^{(N=2)(p)}&={(-q^{1\over 2})_{\infty}(-q)_{\infty}
\over (q)_{\infty} (q^{1\over 2})_{\infty}}\sum_{j=-\infty}^{\infty}
\biggl(q^{j^2p-js+{1\over 2}jp}-q^{j^2p-js-{1\over 2}jp+{s\over 2}}\biggr)\cr
&=q^{-{1\over 4}(s-1)(s-2)}\sum_{n=0}^{\infty}{(-q^{1\over 2})_{n+1} (-q)_n
\over (q)_{2n+1}}q^{n^2+{n\over 2}}F_{1,s}^{(2n+1,p)}(q^{-1}).\cr}}

To calculate the explicit fermionic form in the different sectors
let us first introduce the following matrix
\eqn\CD{C_{D_p}=\left(\matrix{\matrix{2&0\cr0&2\cr}&\matrix{-1&0&\ldots\cr
-1&0&\ldots\cr}\cr\matrix{-1&-1\cr 0&0\cr \vdots&\vdots\cr}&C_{p-2}\cr}\right)}
where $C_{p-2}$ is the $(p-2)$ dimensional Cartan matrix of $A_{p-2}$.
$C_{D_p}$ is the Cartan matrix of the Lie algebra $D_{p}$.

Let us start with the $A$ sector and define $m_0=2n+\rh-\sh$. Then~\ntwoc~
becomes
\eqn\chifermiA{\eqalign{&\chi_{\rh,\sh}^{(N=2)(p)}=
q^{-(\rh-{1\over 2})(\sh-{1\over 2})}\sum_{m_0=0,{\rm restr.}}^{\infty}
{1\over (q)_{m_0}}q^{{1\over 4}(m_0-\rh+\sh)(m_0+\rh-\sh)}
F_{1,\rh+\sh}^{(m_0,p)}(q^{-1})
\cr&~~~~~\times\biggl(
\sum_{k_1=0}^{\infty}y^{{m_0-\rh+\sh\over 2}-k_1}q^{{1\over 2}({m_0-\rh+\sh
\over 2}-k_1)^2}{{m_0-\rh+\sh\over 2} \atopwithdelims[] k_1}_q\biggr)\cr
&~~~~~~\times\biggl(
\sum_{k_2=0}^{\infty}y^{-({m_0+\rh-\sh\over 2}-k_2)}q^{{1\over 2}({m_0+\rh
-\sh\over 2}-k_2)^2}{{m_0+\rh-\sh\over 2} \atopwithdelims[] k_2}_q\biggr)\cr}}
where we used~\isingl~twice to convert $(-yq^{1\over 2})_n$ and $(-y^{-1}
q^{1\over 2})_{n+\rh-\sh}$ into the $k_1$ and $k_2$ sums, respectively. The
restriction on $m_0$ is such that $m_0$ is even if $\rh+\sh$ odd and vice versa
(remember that $\rh,\sh$ are half-integers in the $A$ sector). Then using
{}~\Finv~and setting $\vec{n}=(k_1,k_2,m_0,\vec{m})$ we finally obtain
\eqn\chifermiAa{\eqalign{\chi_{\rh,\sh}^{(N=2)(p)}&=
q^{{1\over 4}(\rh^2+\sh^2-2\rh\sh-\rh-\sh+1)}\sum_{n_1=0}^{\infty}
\sum_{n_2=0}^{\infty}
\sum_{n_i\equiv (\vec{Q}_{\rh,\sh}^{(p)})_i,3\leq i\leq p}
y^{\sh-\rh+n_2-n_1}\cr
&\times q^{{1\over 4}\vec{n}C_{D_p}\vec{n}+{1\over 2}(n_1-n_2)(\rh-\sh)}
{1\over (q)_{n_3}} \prod_{i=1,i\not= 3}^{p}
{{1\over 2}(I_{D_p}\vec{n}+\vec{u}_{\rh,\sh}^{(p)})_i
\atopwithdelims[] n_i}_q\cr}}
where
\eqn\restr{\eqalign{&\vec{Q}_{\rh,\sh}^{(p)}
=\vec{e}_{\rh+\sh+1}+\vec{e}_{\rh+\sh-1}+\ldots\cr
&\vec{u}_{\rh,\sh}^{(p)}=\vec{e}_{\rh+\sh+2}+(\sh-\rh)\vec{e}_1
+(\rh-\sh)\vec{e}_2.\cr}}
Here $\vec{e}_i$ are the $p$ dimensional unit vectors in direction $i$ for
$1\leq i\leq p$.

Similarly we get the fermionic character expression for the $P$ sector from
{}~\ntwpPa~with $m_0=2n+\rh-\sh+1$ and $\vec{n}=(k_1,k_2,m_0,\vec{m})$
\eqn\chifermiP{\eqalign{\chi_{\rh,\sh}^{(N=2)(p)}&=
q^{{1\over 4}(\rh^2+\sh^2-2\rh\sh-\rh-\sh)}\sum_{n_1=0}^{\infty}
\sum_{n_2=0}^{\infty}
\sum_{n_i\equiv (\vec{Q}_{\rh,\sh}^{(p)})_i,3\leq i\leq p}
y^{\sh-\rh+n_2-n_1}\cr
&\times q^{{1\over 4}\vec{n}C_{D_p}\vec{n}+{1\over 2}(n_1-n_2)(\rh-\sh)}
{1\over (q)_{n_3}}
\prod_{i=1,i\not= 3}^{p}{{1\over 2}(I_{D_p}\vec{n}+\vec{u}_{\rh,\sh}^{(p)})_i
\atopwithdelims[] n_i}_q\cr}}
where
\eqn\restra{\eqalign{&\vec{Q}_{\rh,\sh}^{(p)}=\vec{e}_{\rh+\sh+1}
+\vec{e}_{\rh+\sh-1}+\ldots\cr
&\vec{u}_{\rh,\sh}^{(p)}=\vec{e}_{\rh+\sh+2}+(\sh-\rh-1)\vec{e}_1
+(\rh-\sh-1)\vec{e}_2.\cr}}

For the $T$ sector we get from~\Tb~and~\Td~using~\isingl~and~\Finv
\eqn\chifermiT{\chi_s^{(N=2)(p)}=
\sum_{n_1=0}^{\infty}\sum_{n_2=0}^{\infty}\sum_{n_i\equiv(\vec{Q}_s^{(p)})_i,
3\leq i\leq p}q^{{1\over 4}\vec{n}C_{D_p}\vec{n}-{1\over 2}n_1}
{1\over (q)_{n_3}}\prod_{i=1,i\not =3}^{p}{{1\over 2}(I_{D_p}\vec{n}+
\vec{u}_{s}^{(p)})_i\atopwithdelims[] n_i}_q}
where
\eqn\restrT{\eqalign{&\vec{Q}_s^{(p)}=\vec{e}_{s+1}+\vec{e}_{s-1}+\ldots\cr
&\vec{u}_s^{(p)}=\vec{e}_{s+2}+\cases{\vec{e}_1-\vec{e}_2 &for $s$ even\cr
\vec{0} & for $s$ odd\cr}.\cr}}

\newsec{Conclusion}

We have now demonstrated in detail that the construction of Bailey
when applied to the minimal model $M(p,p+1)$ leads to the $N=1$
supersymmetric model $SM(p,p+2)$ and the $N=2$ unitary supersymmetric
model with $c=3(1-2/p).$ However, this construction is merely
illustrative and the method is of great generality. In particular we
can show~\rbms~that if we start from the dual Bailey pair constructed
from the general model $M(p,p')$ (with $p<p'$)we obtain the minimal models
$M(p',2p'-p),$ $SM(p',3p'-2p)$ and the nonunitary $N=2$ models with
$c=3(2p-p')/p'$ while from the direct Bailey pair of $M(p,p')$ we have
the sequence $M(p',p'+p),$ $SM(p',p'+2p)$ and $N=2$ with $c=-3(2p-p')/p'.$
Our results are obtained from the general bosonic form of
Forrester and Baxter~\rfb~and the fermionic results of ~\rbm.

In the unitary case presented in this paper the fermionic
expressions~\chifermiAa, ~\chifermiP~and ~\chifermiT~for the
characters in terms of the Cartan matrix of the group $D_p$ may be
interpreted in terms of the construction of Zamolodchikov and
Fateev~\rzamfat~ in terms of parafermions and two Majorana fermions
if we identify the two variables on the forks of the Dynkin diagram
with the Majorana fermions and the rest of the diagram with the
parafermions $M(1,p+1)$ (which are dual to $M(p,p+1).$)

The general nonunitary
case the $N=2$ are to be compared with the string theory
results of ~\rgrs--\rblnwb~and the flows of minimal models $M(p,p')$
to $M(p',2p'-p)$ is that of ~\rahn.
However, the identification of these character expressions in
the general nonunitary case is more cumbersome than what was done here
in the unitary case and will be treated separately~\rbms.
The flow $M(p,p')$ to $M(p',p'+p)$ does not seem to have been
previously seen.

We also note that the existing computations for the quantum gravity
models are only made explicit for the $W_2$ and $W_3$ gravities. The
$W_2$ case is what is treated here and the $W_3$ case should correspond
to the Bailey pairs of Milne and Lilly~\rml~ constructed from
$SU(3).$ However, in~\rml~the Bailey lemma is derived for all the Lie
groups $A_n$ and $C_n$ and this should correspond to results for all
the $W_n$ gravities.

But probably the most provocative question is to find an
interpretation not only for the Bailey formula for the specialized
values of $\rho_1$ and $\rho_2$ but for general values of these
parameters. These parameters seem to be playing  the role of
fugacities and the Bailey pairs seem to be building up complex systems
by gluing these more elementary fermions together. Thus the parameters
would seem to govern a renormalization flow between models. All of
these questions deserve further study.

\bigskip
{\bf Acknowledgements}

The authors are pleased to acknowledge many useful conversations with
G.Andrews,  J. de Boer,V. Dobrev, O. Foda,W. Nahm, M.Roesgen,
S.O. Warnaar, and N. Warner. This work
is supported in part by the NSF under DMR9404747.
\vfill
\eject
\listrefs

\vfill\eject

\bye
\end